\documentclass[12pt]{article}
\usepackage{amsmath}
\usepackage{amsthm}
\usepackage{enumitem}
\usepackage{afterpage}
\usepackage[autostyle]{csquotes}
\usepackage{mathtools}
\usepackage{cancel}
\usepackage[latin1]{inputenc}
\usepackage{amsmath}
\usepackage{amsfonts}
\usepackage{adjustbox}
\usepackage{amssymb}
\usepackage{framed}
\usepackage{bbm}
\usepackage{hyperref}
\usepackage{rotating} 
\usepackage{xcolor}
\usepackage{graphicx}
\usepackage{dsfont}
\usepackage{bigints}
\usepackage{setspace}
\usepackage{natbib}
\usepackage{booktabs}
\usepackage{multirow}
\usepackage{longtable} 
\usepackage{booktabs}
\usepackage{siunitx}
\usepackage{rotating}

\usepackage[a4paper, total={6.5in, 9in}]{geometry}

\usepackage{changes}

\newtheorem{prop}{Proposition}
\newtheorem{lem}[prop]{Lemma}
\newtheorem{corollary}{Corollary}

\newtheorem{prop1}{Proposition}[section]
\newtheorem{theorem}[prop1]{Theorem}

\newtheorem{rmk1}{Remark}

\newcommand{\ba}{\begin{eqnarray}}
\newcommand{\ea}{\end{eqnarray}}
\newcommand{\bi}{\begin{itemize}}
\newcommand{\ei}{\end{itemize}}
\newcommand{\ben}{\begin{enumerate}}
\newcommand{\een}{\end{enumerate}}
\newcommand{\blem}{\begin{lem}}
\newcommand{\elem}{\end{lem}}
\newcommand{\bteo}{\begin{theorem}}
\newcommand{\eteo}{\end{theorem}}
\newcommand{\bcor}{\begin{corollary}}
\newcommand{\ecor}{\end{corollary}}

\newcommand{\be}{\begin{equation}}
\newcommand{\ee}{\end{equation}}

\title{\textbf{A tale of two sentiment scales: Disentangling short-run and long-run components in multivariate sentiment dynamics}\\
\author{Danilo Vassallo\thanks{%
Scuola Normale Superiore, Italy. E-mail: danilo.vassallo@sns.it} \and Giacomo Bormetti\thanks{%
University of Bologna, Italy E-mail: giacomo.bormetti@unibo.it} \and Fabrizio Lillo\thanks{%
University of Bologna and Scuola Normale Superiore, Italy E-mail: fabrizio.lillo@unibo.it}}
\date{}
}
\begin{document}
\maketitle
\begin{abstract}
We propose a novel approach to sentiment data filtering for a portfolio of assets. In our framework, a dynamic factor model drives the evolution of the observed sentiment and allows to identify two distinct components: a long-term component, modeled as a random walk, and a short-term component driven by a stationary VAR(1) process. 
Our model encompasses alternative approaches available in literature and can be readily estimated by means of Kalman filtering and expectation maximization. This feature makes it convenient when the cross-sectional dimension of the portfolio increases. By applying the model to a portfolio of Dow Jones stocks, we find that the long term component co-integrates with the market principal factor, while the short term one captures transient swings of the market associated with the idiosyncratic components and the correlation structure of returns.  Using quantile regressions, we assess the significance of the contemporaneous and lagged explanatory power of sentiment on returns finding strong statistical evidence when extreme returns, especially negative ones, are considered. Finally, the lagged relation is exploited in a portfolio allocation exercise.\\

\end{abstract}
\vspace{-1.5cm}
\numberwithin{equation}{section}
\section{Introduction}\label{section:introduction}
\noindent Nowadays, as Ignacio Ramonet wrote in The Tyranny of Communication, ``a single copy of the Sunday edition of the New York Times contains more information than an educated person in the eighteenth century would consume in a lifetime". This huge amount of information cannot be read by a single person. Recent developments in machine learning algorithms for sentiment analysis help us to categorise and extract signals from text data and pave the way for a new area of research. The use of these new sources of textual data has become popular to analyse the relationship between sentiment and other economic variables using econometric techniques. \citep{algabaeconometrics} refer to this new strand of literature as \textit{Sentometrics}. For instance, \citep{gross2011machines} study the impact of unexpected news on the displayed quotes in a limit order book, \citep{SUN2016147} show that intraday S\&P 500 index returns are predictable using lagged half-hour
investor sentiment, \citep{Antweiler, Borovkova2015, Allen2015, Smales2015} study the impact of sentiment on volatility, \citep{peterson2016trading} investigates the trading strategies based on sentiment, \citep{Tetlock2007, Garcia2013} consider the Dow Jones Industrial Average (DJIA) index predictability using sentiment, \citep{calomiris2019news} show how the predictability can be exploited in different markets around the world, \citep{Ranco2015} analyse the impact of social media attention on market dynamics, \citep{Borovkova2015_2} develops risk measures based on sentiment index, and \citep{Piilo} show that different types of investors react differently to news sentiment. 

The approaches to sentiment analysis can be broadly classified into three categories. The first class is based on (mostly supervised) Machine Learning techniques. Three steps are typically considered. The first one is to collect textual data forming the training dataset. The second one is to select the text features for classification and to pre-process the data according to the selection. The final step is to apply a classification algorithm to the textual data. As an example, \citep{Pang} compare the performance of Naive Bayes, support vector machines, and maximum entropy algorithm to classify positive or negative movie reviews. The second category is the lexicon-based approach. It also typically consists of three steps. The first step is the selection of a dictionary of $N$ words which could be relevant for a specific topic (e.g. the word \textit{great} is considered as a positive word to review a movie). The second one consists in tokenizing the textual data and, for each word in the dictionary, count how many times it appears in the text. This process can be visualized with a vector of length $N$ where the $i$-th element represents the number of times the $i$-th word of the dictionary is mentioned in the text. Finally, a measure takes the vector of length $N$ as an input and gives a quantitative score as an output. One can refer to~\citep{LOUGHRAN_MCDONALD2011} for a relevant example in the financial literature. The third and last approach is a combination of methodologies coming from the first and second approach. For an overview of textual data treatments and computational techniques, we refer to the review paper \citep{vohra2013comparative} and the book \citep{liu2015sentiment}. 

However, as observed by Zygmunt Bauman in Consuming Life, as the number of information increases also the number of useless information increases, and the noise becomes predominant.
Two different non-exclusive methods have been explored in the literature to remove or, at least, mitigate the impact of useless information. In the first case, a general-to-specific approach is used directly on the textual data. The amount of information can be reduced selecting only verified news (i.e. eliminating fake news), considering only the words which are closely related to the topic of interest, considering the importance of any news (e.g. \citealt{DA2011}), selecting only news which appear for the first time (e.g. Thomson Reuters News Analytics engine uses the novelty variable, see~\citealt{borovkova2017sensr} ), or weighting a news by means of a measure of attention (e.g. with the number of clicks it receives when published in a news portal~\citealt{Ranco2016}). Obviously, the selection of the relevant data is application-specific. For instance, fake news may be irrelevant to forecast the GDP of a country but may be crucial to forecast the results of an election (e.g. \citealt{allcott2017social}).\\
In the second case, sentiment time series are directly considered, rather than the text source they are built from. The observed sentiment is noisy and various approaches have been proposed to filter it and to recover the latent signal. \citep{thorsrud2018words} applies a 60-day moving average, \citep{peterson2016trading} uses the Moving Average Convergence-Divergence methodology proposed in \citep{Appel2003} and \citep{Borovkova2015, Audrino2017, borovkova2017sensr} introduce the Local News Sentiment Level model (LNSL), a univariate method which takes inspiration from the Local Level model of \citep{durbin2012time}.
In spite of its convenience from a practical perspective, the moving average approach is not statistically sound and the window length is usually chosen following rules of thumb, which have been tested empirically but lack a clear theoretical motivation. The methods based on the Kalman-Filter techniques present a natural and computationally simple choice to extract informative signal. Unfortunately, when multiple assets are considered in the analysis, the LNSL model does not exploit the multivariate nature of the data. One goal of this paper is to show that the covariance structure is very informative in sentiment time series analysis.

The first contribution of this paper is to extend the existing time series methods in the latter stream of literature. We propose to model noisy sentiment disentangling two different sentiment signals. In our approach, the observed sentiment follows a linear Gaussian state-space model with three relevant components. The first component, named \textit{long-term sentiment} is modeled as a random walk, the second component is termed \textit{short-term sentiment} and follows a VAR(1) process, and the last component is an i.i.d Gaussian \textit{observation noise} process. We name the novel sentiment state-space model Multivariate Long Short Sentiment (MLSS). We empirically show that the decomposition provides a better insight on the nature of sentiment time series, linking the long-term sentiment to the long-term evolution of the market -- proxied by the market factor -- while the short-term sentiments reflect transient swing of the market mood and is more related to the market idiosyncratic components. Specifically, we find that i) the long-term sentiment cointegrates with the first market factor extracted via PCA; ii) the correlation structure of the short-term sentiment explains a significant and sizable fraction of correlation of return residuals of a CAPM model. Finally, we show that the multivariate local level model provides the best description of the data with respect to alternative models, such as the LNSL. 

The second contribution of the paper is to unravel the relation between news and market returns conditionally on quantile levels. We perform various quantile regressions showing that sentiment has good explanatory power of returns. When contemporaneous effects are considered, the result is expected and holds for all models at intermediate quantile levels. However, when the analysis is focused on abnormal days -- i.e. days for which returns belong to the 1\% and 99\% quantiles -- neither the noisy sentiment nor the filtered sentiment from an LNSL model explain the observed market returns. The only model achieving statistical significance is the MLSS. This result shows that it is essential to filter the noisy sentiment according to the MLSS, which exploits both the multivariate structure of the data and disentangles the long- and short-term components. Moreover, a test performed on the single components confirms the intuition that the short-term sentiment is the one responsible for the contemporaneous explanatory power. The empirical evidence in favor of the MLSS becomes even more compelling when lagged relations are tested. When a single day lag is considered, i.e. one tests whether yesterday sentiment explains today returns, the significance of all models, but MLSS, drops to zero. This result holds across all quantile levels. Instead, for quantiles smaller than 10\% and larger than 90\%, the returns predictability for the MLSS model is highly significant. As before, the decomposition in two time scales is essential and the short-term component is the one responsible of the effect. The analysis extended including lagged sentiment -- up to five days -- confirms previous findings by~\citep{Garcia2013} that past sentiment contributes in predicting present returns. Interestingly, this is true for quantiles between 5\% and 10\%, both negative and positive, but neither in the median region nor for extreme days. In light of this findings, we finally investigated whether media and social news immediately digest market returns and whether this relation depends on the sign of returns. Our results provide a clear picture showing that i) the impact of market returns on sentiment is significant up to five days in the future when negative extreme returns -- i.e. belonging to quantiles from 1\% to 10\% -- are considered, ii) when positive returns are considered the impact rapidly fades out and is significant only for quantiles smaller than 5\%, iii) previous findings become not significant if the MLSS sentiment is replaced by the observed noisy sentiment. Consistently with the intuition provided by these results, we test whether the returns predictability of the MLSS model can be exploited in a portfolio allocation exercise. We show that the portfolios generated with the MLSS sentiment series have higher Sharpe ratio and lower risk than  similar portfolios constructed with raw sentiment or sentiment filtered with the univariate LNSL model. Our model outperforms also the benchmark constituted by the buy-and-hold equally weighted portfolio. This result remains true when transaction costs are included.\\
\indent The rest of the paper is organized as follows. In section \ref{section:model}, we develop the multivariate model for the sentiment and discuss the estimation technique. In section \ref{section:TRMI_index}, we introduce the TRMI sentiment index and describe the data used in the analysis. In section \ref{section:empirical_analysis}, we report the empirical findings and discuss the advantages of the multivariate approach. In section \ref{section:applications}, we compare the various techniques and report the performances of the long-short sentiment decomposition in explaining daily returns. Section \ref{section:portfolio}  describes the portfolio allocation strategies using different filtering techniques and assesses the superiority of the MLSS filter among the others. Section \ref{section:conclusion} draws the relevant conclusions and sketch possible future research directions.
\vspace{-0.5cm}

\section{The Model}\label{section:model}
Consider $K$ assets and the corresponding $K$ observed daily sentiment series $S_t^{i}$ where $i = 1, \ldots, K$. The observed daily sentiment $S_t^{i}$ quantifies the opinions of investors and consumers about company $i$. In most cases, the observed sentiment is a continuous number in a compact set.\\
\indent The Local News Sentiment Level model (LNSL), presented in \citep{Borovkova2015} and subsequently used in \citep{Audrino2017}, reads as follows
\vspace{-0.2cm}
\begin{equation}\label{eq:LNSL}
\begin{aligned}
S^{i}_t &= F^{i}_t + \epsilon_t, \qquad \epsilon_t \overset{d}{\sim} \mathcal{N}\left(0, \sigma^i_{\epsilon} \right),\\
F^{i}_{t} &= F^{i}_{t-1} + v_t, \quad v_t \overset{d}{\sim}  \mathcal{N}\left(0, \sigma^i_{v} \right).
\end{aligned}
\vspace{-0.2cm}
\end{equation}
for every $i = 1,\ldots, K$. This model is a univariate specification of the Local Level model of \citep{durbin2012time}. The latent sentiment series $F^{i}_t$ are considered as slowly changing components, modeled as independent random walks and the parameters $\sigma^{i}_\epsilon$ and $\sigma^{i}_v$ are estimated via maximum likelihood (MLE).\\
\indent Since the LNSL model does not consider the correlations of the innovations among the $K$ assets, we can easily derive its multivariate version as 
\vspace{-0.2cm}
\begin{equation}\label{eq:MLNSL}
\begin{aligned}
S_t &= F_t + \epsilon_t, \qquad \epsilon_t \overset{d}{\sim} \mathcal{N}\left(0, R \right),\\
F_{t} &= F_{t-1} + v_t, \quad v_t \overset{d}{\sim}  \mathcal{N}\left(0, Q \right).
\end{aligned}
\vspace{-0.2cm}
\end{equation}
where $S_t = \left[S^{1}_t, \ldots, S^{K}_t \right]'$ and $F_t = \left[F^{1}_t, \ldots, F^{K}_t \right]'$ are $K$ dimensional vectors, $Q$ is a $K\times K$ symmetric matrix and $R$ is a $K\times K$ diagonal matrix. We refer to the multidimensional LNSL model as MLNSL. The synchronous correlation among the innovations of the latent sentiment are described by the covariance matrix $Q$, while the correlations among the observation noises are assumed to be $0$. Clearly, the LNSL model is a special case of the MLNSL model when the matrix $Q$ is diagonal. Since the number of parameters for this model scales as $K^2$, the MLE of the MLNSL model is computationally demanding. For this reason, we use the Kalman-EM approach described in \citep{KEM}.\\
\indent The idea of the LNSL and MLNSL models is that the latent sentiment is a slowly changing component with a Gaussian disturbance. In their empirical studies, \citep{Audrino2017} observe that the signal to noise ratio $\frac{\sigma_v^2}{\sigma_\epsilon^2}$, obtained using the LNSL filter, is very small. This finding indicates that the majority of the daily changes in the sentiment series can be considered as noise. One possible explanation of this result is that the Local Level specification of these models is not sufficiently rich to capture all the signals from the observed sentiment. Indeed, in newspapers and social media there is a consistent amount of articles and opinions which represent fast trends or rapidly changing consumer preferences. Following the recent strand of literature on persuasion \citep{Gerber_Short_Memory, Seth_Short_Memory}, these fast trends have strong but short-lived effects on consumer preferences. Since the (M)LNSL model interprets the latent sentiment as an integrated series, these signals are considered as noise.\\
\indent The main contribution of this paper is to define a new model which disentangles the slowly changing sentiment from a rapidly changing sentiment, that we name short-term sentiment, and the observation noise. In addition, it is reasonable to think that the slowly changing components of a set of firms with common characteristics, for instance belonging to the same sector, market, or country, should be affected by the same trends and shocks. For this reason, in our model we consider a number $q\leqslant K$ of common factors driving the slow component of the sentiment dynamics. We name these common factors as long-term sentiment. We do not fix the number $q$ \textit{a priori}, but we select it by means of an information criterion.

To provide a more quantitative intuition behind our modeling specification, let us consider the true, but unobserved, daily investor's mood $M^{i}_t$ of asset $i$. We hypothesize that the today daily mood can be written as
\vspace{-0.4cm}
\begin{equation}\label{eq:Mood_change}
\text{Mood}^{i}_{t} = \text{Long-term Mood}^{i}_{t} + \text{Short-term Mood}^{i}_{t}.
\vspace{-0.4cm}
\end{equation}
The Long-term Mood is composed by the yesterday Long-term Mood plus a shock $s^{i,\text{ long}}_t$, which is usually small but permanent, i.e.
\vspace{-0.4cm}
\begin{equation*}
    \text{Long-term Mood}^{i}_{t} = \text{Long-term Mood}^{i}_{t-1} + s^{i,\text{ long}}_t.
    \vspace{-0.4cm}
\end{equation*}
On the contrary, the Short-term Mood is short-lived, but with a strong and highly influential impact. In particular, the Short-term Mood is composed by a residual part of the yesterday Short-term Mood plus a shock $s^{i,\text{ short}}$, i.e.
\vspace{-0.4cm}
\begin{equation*}
    \text{Short-term Mood}^{i}_{t} = \phi^i \text{Short-term Mood}^{i}_{t-1} + s^{i,\text{ short}}_t.
    \vspace{-0.4cm}
\end{equation*}
In this framework, the long-term shocks permanently change the investor's mood while the short-term shocks has an exponentially decaying persistence in the investor's mood. Equation \eqref{eq:Mood_change} can be rewritten as
\vspace{-0.4cm}
\begin{equation}\label{eq:Mood_change_decomposed}
\text{Mood}^{i}_{t} = \text{Long-term Mood}^{i}_{t-1} + s^{i,\text{ long}}_t + \phi^i \text{Short-term Mood}^{i}_{t-1} + s^{i,\text{ short}}_t.
\vspace{-0.4cm}
\end{equation}
Considering the whole story and the dynamic of the two sentiments shocks, we can rewrite equation \eqref{eq:Mood_change_decomposed} as
\vspace{-0.2cm}
\begin{equation*}
\text{Mood}^{i}_{t} =  \underbrace{\sum_{k=-\infty}^{t}(\phi_i)^{t-k} s^{i,\text{ short}}_{k}}_{\text{Short-term Mood}^{i}_{t} } + \underbrace{\sum_{k=-\infty}^{t+1}s^{i,\text{ long}}_{k}}_{\text{Long-term Mood}^{i}_{t} }\,,
\vspace{-0.2cm}
\end{equation*}
where we assumed $\text{Mood}^{i}_{-\infty}$ to be negligible and equal to zero.
In full generality, the multivariate version of model \eqref{eq:Mood_change} can be formulated as follows
\vspace{-0.4cm}
\begin{equation*}
\text{Mood}_{t} = A\, \text{Long-term Mood}_{t} + B\, \text{Short-term Mood}_{t}\,,
\vspace{-0.4cm}
\end{equation*}
with $A$ and $B$ being $K\times K$ matrices. However, in light of the considerations in the previous paragraph, we restrict the matrix $B$ to be the identity matrix. In this way, the Short-term Mood is purely company-specific. We replace $A\, \text{Long-term Mood}_{t}$ with the product between a factor loading matrix and a limited number of long-term and common factors, that is we rewrite the previous equation as 
\vspace{-0.4cm}
\begin{equation}\label{eq:Mood_change_decomposed_multivariate}
\text{Mood}_{t} = \Lambda\, \text{Long-term Factor Mood}_{t} + \text{Short-term Mood}_{t},
\vspace{-0.4cm}
\end{equation}
where $\Lambda$ belongs to $\mathbb{R}^{K\times q}$ with $q\leq K$. It is important to notice that the significance of $\Lambda$  can be statistically tested and the selection of the number $q$ of common factors can be performed by means of AIC and BIC criteria. 
Following \cite{Audrino2017}, we assume that the observed sentiment $S_t$ is a noisy observation of the investors $\text{Mood}_t$, and we formulate a state-space model for $S_t$  consistent with the intuition provided by model \eqref{eq:Mood_change_decomposed_multivariate}. 
The Multivariate Long Short Sentiment model (MLSS) for the observed sentiment model, assuming a Gaussian specification for the short-term sentiment shock, long-term sentiment shock and the observation noise, reads
\vspace{-0.2cm}
\begin{equation}\label{eq:sentiment_model}
\begin{aligned}
S_t &= \Lambda F_t + \Psi_t + \epsilon_t, \quad \epsilon_t \overset{d}{\sim} \mathcal{N}\left(0, R \right),\\
\Psi_t &= \Phi \Psi_{t-1}+u_t, \qquad u_t \overset{d}{\sim} \mathcal{N}\left(0, Q_{short} \right),\\
F_{t} &= F_{t-1} + v_t, \qquad \quad v_t \overset{d}{\sim}  \mathcal{N}\left(0, Q_{long} \right),
\end{aligned}
\vspace{-0.2cm}
\end{equation}
where $R \in \mathbb{R}^{K\times K}$ is the diagonal covariance matrix of the observation noise $\epsilon_t$, $\Phi \in \mathbb{R}^{K\times K}$ is the matrix of autoregressive coefficients, $Q_{short}\in \mathbb{R}^{K\times K}$ is the covariance matrix of the short-term sentiment innovations, and $Q_{long}\in \mathbb{R}^{q\times q}$ is the covariance matrix of the random walk innovations. In equation \eqref{eq:sentiment_model},
$F_t$ and $\Psi_t$ are the latent processes which proxy the Long-term Factor Mood and Short-term Mood in \eqref{eq:Mood_change_decomposed_multivariate}, respectively. 
Please notice that the essential difference between equation \eqref{eq:Mood_change_decomposed_multivariate} and equation \eqref{eq:sentiment_model} is that the observed sentiment, and its components, are noisy versions of the investors' mood and its long and short components. Finally, in this paper, we force a diagonal structure on the matrix $\Phi$, thus neglecting the possible lead-lag effects among sentiments. This restriction is introduced to limit the curse of dimensionality of the model.

\indent The estimation of the unknown parameters is based on a combination of the Kalman filter with Expectation Maximization \citep{KalmanFilter, Shumway1982, WU199699, harvey_1990, Banbura, Jungbacker2008}.
Given that model \eqref{eq:MLNSL} is a special case of model \eqref{eq:sentiment_model}, in Appendix \ref{ap:estimation} of the supplementary material we only consider the estimation procedure of model \eqref{eq:sentiment_model}.
\vspace{-0.5cm}

\section{Data}\label{section:TRMI_index}
\indent The TRMI sentiment index is constructed using over 700 primary sources, divided in news and social media, and collects more than two millions articles per day. For any article, a ``bag-of-words" technique is used to create a sentiment score, which lies between $-1$ and $+1$, a buzz variable\footnote{``The buzz field represents a sum of entity-specific words and phrases used in TRMI computations. It can be non-integer
when any of the words/phrases are described with a minimizer, which reduces the intensity of the primary word or
phrase. For example, in the phrase less concerned the score of the word concerned is minimized by ``less".
Additionally, common words such as ``new" may have a minor but significant contribution to the Innovation TRMI. As a
result, the scores of common words/phrases with minor TRMI contributions can be minimized." See TRMI user guide.}, and one or more asset codes, which in our case refer to companies. The time resolution of the sentiment data is one minute.\\
\indent For any asset $a$, minute $s$, and day $t$ we denote as $S^{a}_{t,s}$ the sentiment score and as $\text{Buzz}^{a}_{t,s}$ the buzz variable. Since the following empirical analysis are performed using daily data, we need to aggregate the TRMI series on a daily basis. TRMI user guide suggests to use the following equation 
\begin{equation}\label{eq:daily_index}
S_t^{a} = \frac{\sum_{s=\text{sh}^{t-1}}^{\text{sh}^{t}} Buzz_{t,s}^{a} S_{t,s}^{a }}{\sum_{s=\text{sh}^{t-1}}^{\text{sh}^{t}} Buzz_{t,s}^{a} }\in \left[-1,1 \right],
\end{equation}
where $S_t^{a}$ refers to the daily sentiment at day $t$, evaluated on a 24-hour window between the selected hour of day $t-1$ ($\text{sh}^{t-1}$) and the selected hour of day $t$ ($\text{sh}^{t}$). Note that the TRMI server provides a daily frequency sentiment, where they use equation \eqref{eq:daily_index} with sh = 3:30 PM. However, since we want to relate the sentiment series with close to close returns, we construct the daily sentiment series aggregating the high-frequency sentiment according to the trading closing hour of the NYSE (sh = 4:00 PM).  For more details, please refer to \citep{peterson2016trading}.\\
\indent For the empirical analysis, we consider the TRMI sentiment index of 27 out of 30\footnote{We only consider 27 assets because one is missing in the Thomson Reuters dataset and two have an high ratio of missing values at the beginning of the sample.} stocks of the Dow Jones Industrial Average (DJIA) over the period 03/01/2006 -- 29/12/2017. Since the TRMI index divides the news sentiment from the social sentiment, we have a total of $54$ time series. A description of tickers and sectors is available in Appendix \ref{ap:tickers_description} of the supplementary material. Finally, the MLSS model, in its current specification, does not manage missing values in data, while some of the sentiment time series present missing observations. The EM algorithm is naturally designed to handle missing observations. However, since the number of missing values is small\footnote{47 out of 54 sentiment series have less than $1\%$ of missing observations. All the series have a percentage of missing which is smaller than $7.5\%$}, we fill them using the rolling mean over the last 5 days.  
\vspace{-0.5cm} 
   
\section{Empirical analysis}\label{section:empirical_analysis}
In this section, we present the results of the estimation of the MLSS model for the investigated stocks, providing an economic interpretation for the long- and short-term component of the sentiment. In the analyses, we consider separately the case of  news and social sentiment indicator.  

The first quantity to fix is the number $q$ of long-term sentiment factors.  Using the Bayesian information criteria (BIC) we select $q_{\text{news}}=2$ and $q_{\text{social}}=2$.

Table \ref{table:static_news_sentiment} reports the values of $\Phi$ and $\Lambda$ with the estimation errors\footnote{Note that the $\Lambda$ matrices, as discussed in the supplementary material, have the upper triangular submatrix equal to zero.}. Bold values indicate parameters which are significantly different from $0$ with a p-value smaller than $0.05$. We notice that most of the estimated parameters are statistically significant. 

\begin{figure}[htbp]
\hspace*{-1.0cm}
\begin{center}
\includegraphics[scale=0.3]{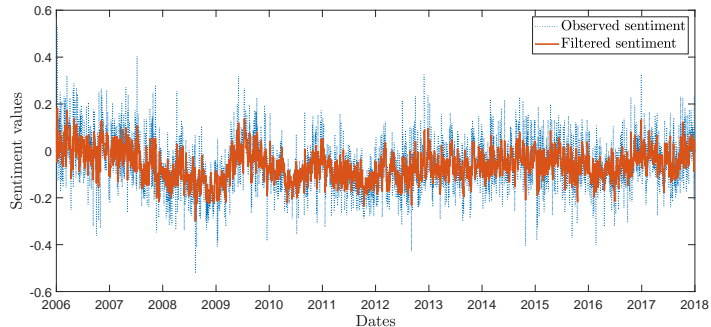} 
\caption{\small Goldman Sachs sentiment series. In blue the observed sentiment, in orange the filtered sentiment including both long-term and short-term component.} \label{fig:example_filter_GS}
\end{center}
\end{figure}   
\indent As an illustrative example, Figure \ref{fig:example_filter_GS} shows how the filter works for the Goldman Sachs news sentiment series. We observe that a high fraction of the sentiment daily variation is captured by the filter. In Appendix \ref{ap:snr} of the supplementary material we quantify more in detail the signal-to-noise ratio of the proposed filter. We find that the MLSS model has a signal-to-noise ratio approximately twenty times larger than the MLNSL. Moreover, the noise in social media is generally higher than the noise in newspapers.
\begin{table}[!htbp]                                                                                                                                                                                                                         
\centering      
\footnotesize
\begin{tabular}{c|cllll}                                                                                                     
\hline
\multirow{2}{*}{\textbf{Tickers}} &  \multirow{2}{*}{\textbf{$\Phi^{\text{news}}$}} & \multicolumn{2}{c}{\multirow{2}{*}{\textbf{$\Lambda^{\text{news}}$}}} & \multicolumn{2}{c}{Signal to noise} \\
& & & & MLSS & MLNSL\\
\hline
AXP & $\bf{0.464}$ & $\bf{1.177}$ &  & $0.623$ & $0.010$ \\                                                           
 & \tiny{($0.029$)} & \tiny{($0.050$)} &  &  &  \\                                                     
JPM & $\bf{0.732}$ & $\bf{-0.169}$ & $\bf{0.711}$ & $0.326$ & $0.023$ \\                                                     
 & \tiny{($0.016$)} & \tiny{($0.035$)} & \tiny{($0.058$)} &  &  \\                                                   
VZ & $\bf{0.682}$ & $\bf{0.545}$ & $-0.080$ & $0.431$ & $0.029$ \\                                                           
 & \tiny{($0.019$)} & \tiny{($0.038$)} & \tiny{($0.063$)} &  &  \\                                                   
CVX & $\bf{0.545}$ & $\bf{0.103}$ & $\bf{0.894}$ & $0.610$ & $0.022$ \\                                                      
 & \tiny{($0.024$)} & \tiny{($0.042$)} & \tiny{($0.071$)} &  &  \\                                                   
GS & $\bf{0.773}$ & $\bf{-0.239}$ & $\bf{0.718}$ & $0.336$ & $0.029$ \\                                                      
 & \tiny{($0.014$)} & \tiny{($0.036$)} & \tiny{($0.060$)} &  &  \\                                                   
JNJ & $\bf{0.407}$ & $\bf{0.851}$ & $\bf{0.834}$ & $0.788$ & $0.010$ \\                                                      
 & \tiny{($0.030$)} & \tiny{($0.039$)} & \tiny{($0.065$)} &  &  \\                                                   
MRK & $\bf{0.336}$ & $\bf{0.811}$ & $\bf{0.885}$ & $0.832$ & $0.008$ \\                                                      
 & \tiny{($0.033$)} & \tiny{($0.036$)} & \tiny{($0.059$)} &  &  \\                                                   
PFE & $\bf{0.299}$ & $\bf{0.530}$ & $\bf{1.021}$ & $1.185$ & $0.007$ \\                                                      
 & \tiny{($0.029$)} & \tiny{($0.031$)} & \tiny{($0.052$)} &  &  \\                                                   
UNH & $\bf{0.374}$ & $\bf{1.177}$ & $\bf{0.530}$ & $0.574$ & $0.009$ \\                                                      
 & \tiny{($0.037$)} & \tiny{($0.056$)} & \tiny{($0.093$)} &  &  \\                                                   
BA & $\bf{0.585}$ & $\bf{0.376}$ & $\bf{0.742}$ & $0.896$ & $0.033$ \\                                                       
 & \tiny{($0.021$)} & \tiny{($0.036$)} & \tiny{($0.059$)} &  &  \\                                                  
CAT & $\bf{0.633}$ & $\bf{0.309}$ & $0.045$ & $0.423$ & $0.017$ \\                                                           
 & \tiny{($0.021$)} & \tiny{($0.064$)} & \tiny{($0.108$)} &  &  \\                                                  
GE & $\bf{0.581}$ & $\bf{1.083}$ & $\bf{-0.196}$ & $0.587$ & $0.022$ \\                                                      
 & \tiny{($0.023$)} & \tiny{($0.035$)} & \tiny{($0.058$)} &  &  \\                                                  
MMM & $\bf{0.295}$ & $\bf{0.958}$ & $0.072$ & $0.788$ & $0.009$ \\                                                           
 & \tiny{($0.034$)} & \tiny{($0.038$)} & \tiny{($0.064$)} &  &  \\                                                  
UTX & $\bf{0.331}$ & $\bf{0.422}$ & $\bf{-0.413}$ & $0.690$ & $0.011$ \\                                                     
 & \tiny{($0.035$)} & \tiny{($0.057$)} & \tiny{($0.094$)} &  &  \\                                                  
XOM & $\bf{0.591}$ & $-0.058$ & $\bf{1.025}$ & $0.725$ & $0.031$ \\                                                          
 & \tiny{($0.021$)} & \tiny{($0.039$)} & \tiny{($0.065$)} &  &  \\                                                  
KO & $\bf{0.486}$ & $\bf{0.476}$ & $\bf{0.245}$ & $0.620$ & $0.015$ \\                                                       
 & \tiny{($0.028$)} & \tiny{($0.033$)} & \tiny{($0.055$)} &  &  \\                                                  
PG & $\bf{0.337}$ & $\bf{0.838}$ & $\bf{-0.623}$ & $0.929$ & $0.008$ \\                                                      
 & \tiny{($0.031$)} & \tiny{($0.041$)} & \tiny{($0.068$)} &  &  \\                                                  
AAPL & $\bf{0.593}$ & $\bf{0.221}$ & $\bf{0.160}$ & $1.736$ & $0.096$ \\                                                     
 & \tiny{($0.018$)} & \tiny{($0.026$)} & \tiny{($0.043$)} &  &  \\                                                  
CSCO & $\bf{0.714}$ & $\bf{1.063}$ & $\bf{-1.094}$ & $0.441$ & $0.046$ \\                                                    
 & \tiny{($0.017$)} & \tiny{($0.043$)} & \tiny{($0.071$)} &  &  \\                                                  
IBM & $\bf{0.603}$ & $\bf{0.754}$ & $\bf{-1.269}$ & $0.853$ & $0.040$ \\                                                     
 & \tiny{($0.020$)} & \tiny{($0.038$)} & \tiny{($0.063$)} &  &  \\                                                  
INTC & $\bf{0.641}$ & $\bf{0.641}$ & $\bf{-0.299}$ & $0.865$ & $0.066$ \\                                                    
 & \tiny{($0.018$)} & \tiny{($0.039$)} & \tiny{($0.065$)} &  &  \\                                                  
MSFT & $\bf{0.651}$ & $\bf{0.858}$ & $-0.007$ & $0.668$ & $0.053$ \\                                                         
 & \tiny{($0.019$)} & \tiny{($0.026$)} & \tiny{($0.043$)} &  &  \\                                                  
DIS & $\bf{0.439}$ & $\bf{0.454}$ & $\bf{-0.198}$ & $1.074$ & $0.013$ \\                                                     
 & \tiny{($0.025$)} & \tiny{($0.028$)} & \tiny{($0.046$)} &  &  \\                                                  
HD & $\bf{0.611}$ & $\bf{1.137}$ & $\bf{0.232}$ & $0.473$ & $0.021$ \\                                                       
 & \tiny{($0.024$)} & \tiny{($0.058$)} & \tiny{($0.098$)} &  &  \\                                                  
MCD & $\bf{0.404}$ & $\bf{-0.291}$ & $0.020$ & $1.401$ & $0.013$ \\                                                          
 & \tiny{($0.024$)} & \tiny{($0.034$)} & \tiny{($0.057$)} &  &  \\                                                  
NKE & $\bf{0.368}$ & $\bf{0.664}$ & $\bf{-0.285}$ & $0.783$ & $0.010$ \\                                                     
 & \tiny{($0.032$)} & \tiny{($0.046$)} & \tiny{($0.076$)} &  &  \\                                                  
WMT & $\bf{0.516}$ & $\bf{0.147}$ & $\bf{0.619}$ & $0.854$ & $0.022$ \\                                                      
 & \tiny{($0.023$)} & \tiny{($0.031$)} & \tiny{($0.052$)} &  &  \\
\hline                                                                                                                                                                                                                                    
\end{tabular}                                                                                                                                                                                                                              
\caption{\small Static parameters of model \eqref{eq:sentiment_model} for news sentiment. Values and standard errors of estimated $\Lambda$ are multiplied by $10^3$. In parenthesis we show the standard error of the estimated parameter. The last two columns show the signal to noise ratio for two competing models.}
\label{table:static_news_sentiment}
\end{table}     

\indent The MLSS approach considers two new quantities extracted from the observed sentiment. The first novelty is the long-term sentiment which, by construction, represents the series of common trends in a particular basket of sentiment time series. The second novelty is the multivariate structure of sentiment, extracted using the symmetric matrix $Q_{short}$. In the next sections, we separately analyse the relation between these two quantities and the stock market prices.
To this end, we extract the market factors from the stock prices of these assets. Denote as $r_t \in \mathbb{R}^{27}$ the vector of demeaned close-to-close log-returns and evaluate the unconditional covariance matrix $Q_{\text{ret}}$ and the unconditional correlation matrix $C_{\text{ret}}$. We extract the factor loading matrix $\Lambda^{\text{mrk}}\in \mathbb{R}^{q_{\text{mrk}} \times 27}$ using the PCA on the matrix $C_{\text{ret}}$ and define the return factors $R_t = \Lambda^{\text{mrk}}r_t\in \mathbb{R}^{q_{\text{mrk}}}$. We also define the market factors as $M_t^{\text{mrk}} = \Lambda^{\text{mrk}}p_t$, where $p_t\in\mathbb{R}^{27}$ is the vector of log-prices. In the following analysis, we consider $q_{\text{mrk}}=1$ and name the first market factor Dow 27.
\vspace{-0.5cm}

\subsection{Long-term Sentiment}\label{sec:long_term_sentiment}

\begin{figure}[htbp]
\hspace*{-1.0cm}
\begin{center}
\includegraphics[scale=0.3]{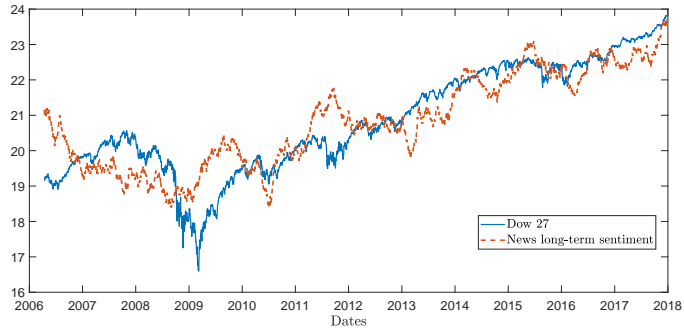} 
\caption{\small Co-integration between Dow 27, in blue, and the second factor of the news long-term sentiment, in orange. Time series are scaled.} \label{fig:news_coint_Granger}
\end{center}
\end{figure}

\begin{figure}[htbp]
\hspace*{-1.0cm}
\begin{center}
\includegraphics[scale=0.3]{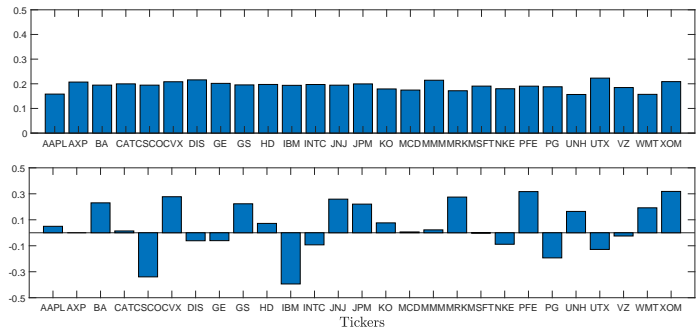} 
\caption{\small Values of the standardized factor loadings of the cointegrated series. Top panel: loadings of the Dow 27 index. Bottom panel: loadings of the second factor of the news long-term sentiment.} \label{fig:news_coint_Granger_loadings}
\end{center}
\end{figure}

We first investigate the economic meaning of the long-term sentiment. Using the Engle-Granger test \citep{Engle_Granger}, we observe that one of the factors of the long-term sentiment is cointegrated with the Dow 27. Figure \ref{fig:news_coint_Granger} shows the cointegration relation, pointing out that the main driver of the prices and the driver of the sentiment time series reflect the same common information. This result per se is not surprising. However, Figure \ref{fig:news_coint_Granger_loadings} shows the standardized weights of the cointegrated factors. The weights of the market factor are very homogeneous across assets, as shown in the top panel, while the weights of the cointegrated factor of the long-term sentiment are very heterogeneous, as shown in the bottom panel. The values of the elements of the factor loading matrix $\Lambda^{\text{news}}$ reported in Table \ref{table:static_news_sentiment} are either positive or negative~\footnote{The elements of the factor loading matrix $\Lambda^{\text{social}}$ are available upon request.}. Then, some firm's sentiment positively affects the common sentiment factors, while some other firm's sentiment negatively affects them. We checked whether the heterogeneity of weights were related with the number of news of a given asset, or with the buzz index, but we found no significant evidence. Unravelling the origin of the detected heterogeneity is an interesting research question, that could be probably answered by looking at the contents of the articles from which the sentiment was computed. Unfortunately, we do not have access to this kind of information.
\vspace{-0.5cm}

\subsection{Short-term Sentiment}\label{sec:short_term_sentiment}
\indent The second novelty of the MLSS model is the multivariate structure of the short-term sentiment series. The question we want to address in this section is whether the correlation structure of the short-term sentiment is (linearly) related with the correlation structure of the daily returns. In the previous section, we observed that one of the factors of the long-term sentiment is cointegrated with the first market factor. We therefore expect the short-term sentiment to capture asset-specific features, i.e. we expect a close relation with the idiosyncratic dynamic of the returns\footnote{We define idiosyncratic returns as the market returns where the first market factor is removed using the factor model \eqref{CAPM}}. To test this intuition for the correlation structure, we compare the results of the MLSS model with the results of the MLNSL model which, by construction, does not disentangle the factors from the sentiment series. If the intuition is correct, the correlation matrix of the sentiment extracted using the MLSS model should be linearly related with the return correlations and with the idiosyncratic return correlations. On the contrary, the correlation matrix of the sentiment extracted using the MLNSL model, which only captures the slowly changing dynamics of the sentiment series, and thus of the first market factor, should be linearly related with the returns correlation but mildly correlated with the idiosyncratic returns correlations. Finally, to test whether the filtering procedure is a crucial step in our approach, the correlation matrix of the observed sentiment is also considered.

We define $C_{\text{short}}$ as the correlation matrix associated with the covariance matrix $Q_{\text{short}}$, $C_{\text{MLNSL}}$ the correlation matrix associated with the covariance matrix $Q$ of equation \eqref{eq:MLNSL}, $C_{\text{Obs}} = \text{Corr}\left(\Delta S_t \right)$ the unconditional correlation of the first difference of the observed sentiment, and $C_{\text{ret}}$ the unconditional correlations matrix of the stock returns. We search for a linear element-wise relation between $C_{\text{ret}}$ and $C_{\text{model}}$, where ${\text{model}}$ is one of ${\text{short}}$, ${\text{MLNSL}}$, or ${\text{Obs}}$. The results are reported for the news case only, but the conclusions are similar for the social sentiment.\\
\indent We perform a standard ordinary least squares estimation on the model
\vspace{-0.4cm}
\begin{equation}\label{eq:OLS_short_corr}
\textsf{vechl}(C_{\text{ret}}) = \alpha + \beta^{\text{model}} \textsf{vechl}(C_{\text{model}}),
\vspace{-0.4cm}
\end{equation} 
where $\textsf{vechl}(X)$ is the operator which collects the upper diagonal elements of matrix $X$ in a column vector. We compare the results obtained using the MLSS model ($C_{\text{model}} = C_{\text{short}}$), with the results obtained using the MLNSL model ($C_{\text{model}} = C_{\text{MLNSL}}$) and using the Observed sentiment ($C_{\text{model}} = C_{\text{Obs}}$). In addition, since the unconditional correlation between two assets is higher when they belong to the same sector, we separately consider two cases. In the first case, we estimate model \eqref{eq:OLS_short_corr} considering all the pairs of assets. In the second case, we estimate model \eqref{eq:OLS_short_corr} considering only the pairs of assets belonging to the same economic sector according to Table \ref{table:tickers_sectors}. 

\indent The top left panel of Table \ref{table:OLS_correlation_results} shows the results with all the correlation pairs. In the first column we report the $R^2$ of the regression, in the second column we report the F-statistic and the relative p-value is reported in the third column. The regressions with $C_{\text{short}}$ and $C_{\text{MLNSL}}$ have high and significant p-values, while the regression with $C_{\text{obs}}$ is not statistically different from the model with the intercept only. This finding has two implications. The first one is that the sentiment innovations have a similar correlation structure of the returns innovations. In particular, if the returns of two assets are relatively highly correlated, then also the increment of the filtered sentiment of the news about these assets are relatively highly correlated. The second implication is that, if a filtering procedure is not applied on the observed sentiment data, the noise is too large to find significant results. In the top right panel of Table \ref{table:OLS_correlation_results} we report the results of the model \eqref{eq:OLS_short_corr} applied to the pairs of assets belonging to the same sector. We observe that the $R^2$ increases for all models. This result is expected since it is well known that the return correlation is higher and more significant between two assets of the same sector. However, even if the $R^2$ increases, the number of pairs decreases. For this reason, the increment in the $R^2$ does not lead to an increment in the $F$-statistic, which fails to reject the null hypothesis for the $C_{\text{obs}}$. This result confirms that the $C_{\text{obs}}$ matrix is not a significant regressor for $C_{\text{ret}}$.

\begin{table}[t]                                                   
\centering      
{\small                                                  
\begin{tabular}{c|lll|lll}                                  
\hline                     
\multirow{2}{*}{\textbf{Models}} &  \multicolumn{3}{c|}{All assets} & \multicolumn{3}{c}{Same sector} \\                  
 & $R^2$ & $F$-statistic & $p$-value & $R^2$ & $F$-statistic & $p$-value \\    
\hline                                
MLSS & 13.77 \% & 55.713 & 0.0000 & 37.89 \% & 23.182 & 0.0000 \\  
MLNSL & 15.63 \% & 64.669 & 0.0000 & 28.78 \% & 15.359 & 0.0004 \\
Obs & 0.95 \% & 3.330 & 0.0689 & 4.19 \% & 1.662 & 0.2052 \\
\hline  
MLSS & 11.34 \% & 44.659 & 0.0000 & 30.91 \% & 17.001 & 0.0002 \\
MLNSL & 4.31 \% & 15.700 & 0.0001 & 7.50 \% & 3.081 & 0.0873 \\ 
Obs & 1.01 \% & 3.554 & 0.0602 & 4.88 \% & 1.950 & 0.1707 \\  
\hline                                                  
\end{tabular}                                            
\caption{\small Top rows: Results from the linear regression \eqref{eq:OLS_short_corr}. Bottom rows: Results from the linear regression \eqref{eq:OLS_short_corr_res}. Left columns: OLS estimates when all the assets are considered; right columns: OLS estimates when only the correlations between stocks belonging to  the same sector are considered. $\text{Obs}$ rows: estimation based on the observed sentiment. }                                          
\label{table:OLS_correlation_results}  
}                                     
\end{table} 
\indent Comparing the top panels of Table \ref{table:OLS_correlation_results}, we note that the increment in the $R^2$ is higher for the MLSS model rather than the MLNSL model. This evidence is consistent with the intuition that the short-term sentiment series, extracted using the MLSS model, are more related with the idiosyncratic returns. Indeed the correlation induced by the market factor is predominant in the first case, reported in the top left panel, where all the assets are considered, rather than the second case, reported in the top right panel, where the co-movements are not only driven by the first market factor, but they are also driven by sector-specific factors.\\
\indent Now we extract the Dow 27 return from the asset returns using a one-factor model. We repeat the analysis comparing the matrices $C_\text{short}$, $C_\text{MLNSL}$ and $C_\text{Obs}$ with the unconditional correlation of the idiosyncratic returns. We extract the market factor $R_t$ from the returns using the factor model
\vspace{-0.2cm}
\begin{equation}\label{CAPM}
r_t^{i} = \alpha^{i}+ \beta^{i} R_t + z_t^{i}, \quad \forall i = 1,\ldots,27
\vspace{-0.2cm}
\end{equation}
where $z_t^{i}\sim N(0,\tilde{Q}_{\text{ret}})$. We then compute the cross-correlation matrix $\tilde{C}_{\text{ret}}$ from the covariance matrix $\tilde{Q}_{\text{ret}}$ and estimate the following model
\vspace{-0.4cm}
\begin{equation}\label{eq:OLS_short_corr_res}
\textsf{vechl}(\tilde{C}_{\text{ret}}) = \alpha + \beta^{\text{model}} \textsf{vechl}(C_{\text{model}}).
\vspace{-0.4cm}
\end{equation} 
The bottom panels of Table \ref{table:OLS_correlation_results} report the results. In the bottom left panel we show the results for the model \eqref{eq:OLS_short_corr_res} where all the correlation pairs are considered.  The first evidence is that the MLNSL $R^2$ dramatically decreases, while the MLSS $R^2$ remains almost the same. This finding suggests that almost all the return correlations explained by the $C_\text{MLNSL}$ matrix are associated with the market factor $R_t$, while the matrix $C_\text{short}$, which represents the fast trends on the sentiment data, also captures different dynamics.\\
In the bottom right panel, we show the results for the model \eqref{eq:OLS_short_corr_res} where we consider only the correlation pairs for assets belonging to the same sector. In this case the differences between the MLSS and MLNSL are more severe. Indeed, the MLSS model still has a high and highly significant $R^2$, while the $F$-statistic for the MLNSL model fails to reject the null that $\beta^{\text{MLNSL}}$, defined in equation \eqref{eq:OLS_short_corr_res}, is equal to $0$. Again, the model with the observed sentiment has not significant p-values.

\begin{figure}[htbp]
\hspace*{-1.0cm}
\begin{center}
\includegraphics[scale=0.3]{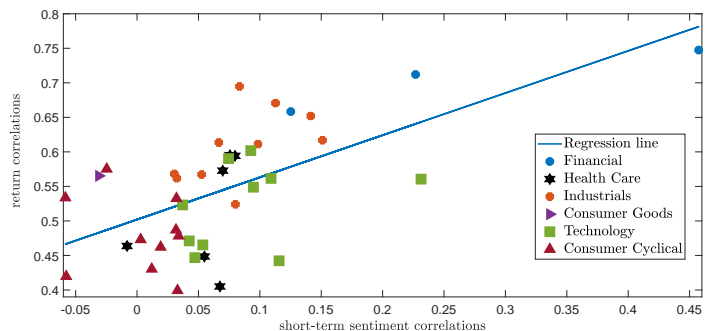} 
\caption{\small Scatter plot of the news short-term sentiment correlations and the return correlations for pairs of assets in the same sector. The line corresponds to the regression \eqref{eq:OLS_short_corr}.} \label{fig:news_corr_same_sector}
\end{center}
\end{figure}

\indent As a last observation, we see the different behavior of the sectors in this regression exercise. 
Figure \ref{fig:news_corr_same_sector} reports the scatter plot of the elements of $C_{\text{short}}$ versus the corresponding values of $C_{\text{ret}}$ when the two stocks belong to the same economic sector, characterized by a specific marker. We also superimpose the regression line obtained from equation \eqref{eq:OLS_short_corr}. Note that the behavior is different among sectors. The financial sector, marked with blue dots, is the one with highest linear relation and the three assets belonging to this sector have all high returns and sentiment correlations. On the contrary, the consumer cyclical sector, marked with garnet-red triangles, has a high dispersion among the correlations of the $5$ assets.

In summary, Sections \ref{sec:long_term_sentiment} and \ref{sec:short_term_sentiment} support the intuition behind the MLSS model. Indeed, the slowly changing components of the sentiment are effectively captured by the long-term sentiment. We successfully confirmed this hypothesis in Section \ref{sec:long_term_sentiment}. At the same time, the short-term sentiment effectively describes the firm-specific behavior of the returns. Section \ref{sec:short_term_sentiment} shows that the MLSS model can capture different features of the returns, while the MLNSL mainly captures the sentiment component associated with the market. 
\vspace{-0.5cm}

\section{Contemporaneous and lagged relations}\label{section:applications}

The goal of this section is to assess the explanatory power of the sentiment with respect to the market returns using the different filters presented in the previous sections. In particular, we show that both the extraction of long-term and short-term sentiment components and the multivariate specification of the model are crucial ingredients to capture the synchronous and lagged effects.

We consider the asset prices $P^i_t$ of the 27 stocks of the Dow 30 and construct the equally weighted portfolio 
\begin{equation}\label{eq:def_market_portfolio}
M_t=\frac{1}{27}\sum_{i=1}^{27} P^{i}_t
\end{equation}
as a representative portfolio and denote with $r^{m}_t$ its log-returns. We consider a representative portfolio for two reasons. Firstly, \citep{beckers2018social} shows that the returns predictability using sentiment indicators is higher when using market indexes rather than single stocks. Secondly, using a representative portfolio we can compare different filtering techniques which do or do not consider the multivariate structure.\\
We define $\bar{S}^{\text{news}}_t = \frac{1}{27}\sum_{i=1}^{27} S^{i,news}_t$ and $\bar{S}^{\text{social}}_t = \frac{1}{27}\sum_{i=1}^{27} S^{i,social}_t$ as the sentiment associated to the representative portfolio. We consider five different filtering techniques defined as follow:
\begin{enumerate}
\item{$S^{MLSS}_t$ is the filtered signal obtained using the MLSS model in equation \eqref{eq:sentiment_model}. The resulting filtered quantities are 4 long-term sentiment factors $F^{MLSS}_t$, 2 for the news and 2 for the social sentiment, and 54 short-term sentiment series $\Psi^{MLSS}_t$, 27 for the news and 27 for the social sentiment. We compute the cross-sectional average for the news short-term sentiment $\bar{\Psi}^{MLSS,news}_t$ and social short-term sentiment $\bar{\Psi}^{MLSS,social}_t$. As a final result, we define
\vspace{-0.4cm}
\[
S^{MLSS}_t = \left[\Delta F_t^{MLSS,news}, \Delta F_t^{MLSS,social}, \bar{\Psi}^{MLSS,news}_t, \bar{\Psi}^{MLSS,social}_t \right]' \in \mathbb{R}^{6}. 
\vspace{-0.4cm}
\] }
\item{$S^{LSS}_t$ is the filtered signal obtained applying the MLSS model directly to the univariate series  $\bar{S}^{\text{news}}_t$ and  $\bar{S}^{\text{social}}_t$. For identifiability reasons, the number of common factors is one. The motivation behind this model is to test whether a simple cross-sectional average of sentiment time series can be an effective proxy of the sentiment of the representative asset. This approach intentionally neglects the multivariate structure of the sentiment and treats it as a non relevant feature. A similar reasoning has been used in \citep{borovkova2017sensr}. The resulting filtered quantities are 2 long-term sentiment factors $F^{LSS}_t$, one for the news and one for the social sentiment, and 2 short-term sentiment series $\bar{\Psi}^{LSS}_t$, one for the news and one for the social sentiment. The final model reads
\vspace{-0.4cm}
\[
S^{LSS}_t = \left[\Delta F_t^{LSS,news}, \Delta F_t^{LSS,social}, \bar{\Psi}^{LSS,news}_t, \bar{\Psi}^{LSS,social}_t \right]' \in \mathbb{R}^{4}. 
\vspace{-0.4cm}
\] }
\item{$S^{MLNSL}_t$ is the filtered signal obtained using the MLNSL model in equation \eqref{eq:MLNSL} from the 54 observed sentiment time series. The resulting filtered quantities are 54 filtered sentiment series $F^{MLNSL}_t$, 27 for the news and 27 for the social sentiment. We compute the cross-sectional average for the news sentiment $\bar{F}^{MLNSL,news}_t$ and social sentiment $\bar{F}^{MLNSL,social}_t$. As a final result, we define
\vspace{-0.4cm}
\[
S^{MLNSL}_t = \left[\Delta \bar{F}^{MLNSL,news}_t, \Delta \bar{F}^{MLNSL,social}_t\right]' \in \mathbb{R}^{2}. 
\vspace{-0.4cm}
\] }
\item{$S^{LNSL}_t$ is the filtered signal obtained applying the LNSL model, introduced by \citep{Borovkova2015} and presented in equation \eqref{eq:LNSL}, to $\bar{S}^{\text{news}}_t$ and $\bar{S}^{\text{social}}_t$. As for the LSS model, the motivation behind this choice is to test whether the multivariate structure of sentiment is a relevant feature or not. We obtain two filtered sentiment series $\bar{F}^{LNSL}_t$, one for the news and one for the social sentiment. We then define 
\vspace{-0.4cm}
\[
S^{LNSL}_t = \left[\Delta \bar{F}^{LNSL,news}_t, \Delta \bar{F}^{LNSL,social}_t\right]' \in \mathbb{R}^{2}. 
\vspace{-0.4cm}
\] }
\item{$S^{obs}_t$ only considers the observed sentiment $\bar{S}^{\text{news}}_t$ and  $\bar{S}^{\text{social}}_t$
\vspace{-0.4cm}
\[
S^{Obs}_t = \left[\Delta \bar{S}^{\text{news}}_t, \Delta \bar{S}^{\text{social}}_t\right]' \in \mathbb{R}^{2}. 
\vspace{-0.4cm}
\] }
\end{enumerate}
In summary, the five models allow us to separate the effect of the different components. The MLSS model exploits all the possible information from the multivariate time series and all the relevant common factors are considered. The average across assets is computed at a later stage on the short-term sentiment. For this reason, it does not affect the long-term components.  The LSS model computes the cross-sectional average as a first step and does not exploit the multivariate structure. Then, both the short-term and long-term components are different from the one of the MLSS model. The MLNSL and LNSL models differ only on the step of the aggregation. The first model applies the filter on the multivariate time series, while the second model applies the filter on the aggregated time series. Finally, the Obs model works as a benchmark.
\vspace{-0.5cm}

\subsection{Quantile regression}
In this section, we investigate the lagged relation between sentiment and market returns. The recent literature for the DJIA \citep{Garcia2013} and for the gold futures \citep{SMALES2014275} found that the reaction to news is more pronounced during recessions. For this reason, we use the quantile regression in place of a simple linear regression to obtain a more comprehensive analysis of the relationship between variables. In Appendix \ref{ap:Contemporaneous_quantile} of the supplementary material we report the investigation on the contemporaneous relation between sentiment and returns.

\subsubsection{Lagged relations}\label{sec:Lagged_quantile}
We consider the following quantile regression
\vspace{-0.4cm}
\begin{equation*}
r^{m}(\tau) = \alpha\left(\tau \right) + \beta^{{\text{model}}}\left( \tau \right) S^{{\text{model}}}_{t-h}\,,
\vspace{-0.4cm}
\end{equation*}
where \textit{model} denotes one of the five filtering models presented above.
According to \citep{Koenker_Machado1999}, we can compare the explanatory power of a selected model according to the $R^1$ measure. In particular, if we consider the functional expression for the quantile regression
\begin{equation}\label{eq:functional_quantile_reg}
\hat{V}\left(\tau\right) =\min_{\left(\alpha, \beta\right)} \sum_{t=1}^{T} \rho_{\tau}\left(r^{m}_t - \alpha - \beta S_{t-h} \right),
\end{equation}
where $\rho_{\tau} (u) = u(\tau-I_{u<0})$, we can define the quantile $R^1$ measure as
\vspace{-0.4cm}
\begin{equation*}
R^1\left(\tau \right) = 1 - \frac{\hat{V}(\tau)}{\tilde{V}(\tau)}\,,
\vspace{-0.4cm}
\end{equation*}
where $\tilde{V}(\tau)$ is evaluated restricting equation \eqref{eq:functional_quantile_reg} with the intercept parameter only.
In contrast with the $R^2$ measure of the linear models, $R^1(\tau)$ is a local measure of goodness of fit and only applies to a particular quantile. In addition, \citep{Koenker_Machado1999} show that using $\hat{V}$ we can test the significance of the $\beta^{\text{model}}$ parameters. Considering $\beta^{\text{model}}=0 $ as the null hypothesis and $F$ as the probability distribution of the i.i.d. residuals $\{u_i\}$, the statistic
\begin{equation}\label{eq:test_quantile_reg}
L_T (\tau) = \frac{2(\tilde{V}(\tau) - \hat{V}(\tau) )}{\tau(1-\tau) s(\tau) } \to \chi^2_{q}
\end{equation}
where $q$ is the dimension of $\beta^{\text{model}}$ and $s(\tau) = 1/f(F^{-1}(\tau))$.\\
\indent As a first step, we consider $h=1$. We evaluate the $R^1(\tau)$ statistic and test the significance using the $\chi^2$-test.
\begin{table}                                                                                 
\centering           
{\small                                                                         
\begin{tabular}{clllll}
\hline                                                                                        
\multirow{2}{*}{\bf{$\tau$ quantiles}} & \multicolumn{5}{c}{\bf{$R^1(\tau)$ measure}}  \\
 & MLSS & LSS & MLNSL & LNSL & Obs \\                                                             
\hline                                                                                        
$0.01$ & $12.7 \%^{***}$ & $4.5 \%$ & $0.3 \%$ & $0.2 \%$ & $0.1 \%$ \\    
$0.05$ & $3.2 \%^{***}$ & $1.3 \%^{**}$ & $0.1 \%$ & $0.0 \%$ & $0.1 \%$ \\
$0.10$ & $1.7 \%^{***}$ & $1.2 \%^{***}$ & $0.0 \%$ & $0.0 \%$ & $0.1 \%$ \\
$0.33$ & $0.2 \%$ & $0.1 \%$ & $0.0 \%$ & $0.0 \%$ & $0.0 \%$ \\            
$0.50$ & $0.2 \%^{*}$ & $0.1 \%$ & $0.1 \%$ & $0.1 \%$ & $0.0 \%$ \\        
$0.66$ & $0.4 \%^{**}$ & $0.2 \%$ & $0.1 \%$ & $0.1 \%$ & $0.0 \%$ \\      
$0.90$ & $2.8 \%^{***}$ & $1.0 \%^{***}$ & $0.2 \%$ & $0.1 \%$ & $0.1 \%$ \\
$0.95$ & $5.3 \%^{***}$ & $1.6 \%^{***}$ & $0.3 \%$ & $0.1 \%$ & $0.2 \%$ \\
$0.99$ & $11.9 \%^{***}$ & $3.4 \%$ & $0.0 \%$ & $0.5 \%$ & $1.0 \%$ \\ 
\hline                                                                                        
\end{tabular}                                                                                 
\caption{\small The $R^1$ measure across the value $\tau$ for the one-lag quantile regression. We denote with $^{***}$ the significance at $1\%$, $^{**}$ the significance at $5\%$ and $^{*}$ the significance at $10\%$ }                                                                      
\label{table:R1_quantile_reg_lagged}                          
}                                          
\end{table}
Table~\ref{table:R1_quantile_reg_lagged} reports the values and significance of the $R^1$ measure. A finding is common among all models: the values of $R^1$ are higher in the tails and lower close to the median. In addition, what we observe is extremely promising for the Long-Short modeling approach. The significance of the noisy sentiment is zero for all quantile levels. Filtering the time series is essential to recover predictability. However, filtering alone is not sufficient. Indeed, neither the predictability  of the LSNL model nor of the multivariate extension MLSNL is statistically significant. Significance is recovered only when the filtered sentiment is decomposed into the short-run and long-run components. This is true for extreme returns, both positive and negative. The result is stronger when the LSS model is replaced by the MLSS, meaning that the cross-sectional dependence is an important ingredient to enhance predictability.

A further advantage of the long-short decomposition is that we can properly asses the relative contribution of the two components. In particular, we use equation \eqref{eq:test_quantile_reg} to test the significance of the parameters in the MLSS model.
Considering the $S^{MLSS} = \left[\Delta F_t^{MLSS}, \bar{\Psi}^{MLSS}_t \right]$, the significance of the parameter $\beta^{LT} \in \mathbb{R}^{4}$ and $\beta^{ST} \in \mathbb{R}^{2}$ can be tested using
\vspace{-0.4cm}
\begin{equation*}
\tilde{V}^{LT}\left(\tau\right) =\min_{\left(\alpha, \beta^{LT}\right)} \sum_{t=1}^{T} \rho_{\tau}\left(r^{m}_t - \alpha - \beta^{LT} \Delta F_{t-h}^{MLSS} \right)
\vspace{-0.4cm}
\end{equation*}
and
\begin{equation*}
\tilde{V}^{ST}\left(\tau\right) =\min_{\left(\alpha, \beta^{ST}\right)} \sum_{t=1}^{T} \rho_{\tau}\left(r^{m}_t - \alpha - \beta^{ST} \bar{\Psi}^{MLSS}_{t-h} \right),
\end{equation*}
which lead to the statistics
\vspace{-0.4cm}
\begin{equation}\label{eq:statistic_LT_quant_reg}
L^{LT} (\tau) = \frac{2( \tilde{V}^{ST}(\tau) - \hat{V}(\tau) )}{\tau(1-\tau) s(\tau) } \to \chi^2_{4}
\vspace{-0.4cm}
\end{equation}
and
\begin{equation}\label{eq:statistic_ST_quant_reg}
L^{ST} (\tau) = \frac{2( \tilde{V}^{LT}(\tau) - \hat{V}(\tau) )}{\tau(1-\tau) s(\tau) } \to \chi^2_{2}\,.
\end{equation}   
We report the p-values of the test statistics \eqref{eq:statistic_LT_quant_reg} and \eqref{eq:statistic_ST_quant_reg} in Table~\ref{table:p_val_MLSS_quant_reg_lagged}.
\begin{table}                       
\centering            
{\small              
\begin{tabular}{cll}            
\hline                              
\multirow{2}{*}{\bf{$\tau$ quantiles}} & \multicolumn{2}{c}{\bf{$p$-values}}             \\       
 & $L_{t-1}^{ST}$ &  $L_{t-1}^{LT}$ \\             
\hline                              
$0.01$ & $0.020 \%$ & $67.688 \%$ \\
$0.05$ & $0.000 \%$ & $66.738 \%$ \\
$0.10$ & $0.003 \%$ & $73.881 \%$ \\
$0.33$ & $5.069 \%$ & $66.668 \%$ \\
$0.50$ & $16.360 \%$ & $59.465 \%$ \\
$0.66$ & $7.692 \%$ & $7.411 \%$ \\  
$0.90$ & $0.000 \%$ & $0.011 \%$ \\  
$0.95$ & $0.000 \%$ & $1.789 \%$ \\  
$0.99$ & $0.001 \%$ & $57.969 \%$ \\ 
\hline                              
\end{tabular}                       
\caption{\small p-values for the statistics $L_{t-1}^{ST}\sim \chi^2_{2}$ and $L_{t-1}^{LT}\sim \chi^2_{4}$ defined in a similar fashion to equations~\eqref{eq:statistic_ST_quant_reg} and~\eqref{eq:statistic_LT_quant_reg}.}
\label{table:p_val_MLSS_quant_reg_lagged}   
}       
\end{table}   
The contribution given by the short-term sentiment is strongly significant, in particular for extreme quantiles. On the contrary, the long-term sentiment is not significant in 6 out of 9 quantiles. The results support the intuition that, if today a very high or very low return appears, it can be partially explained by the yesterday's rapidly changing mood, while the permanent trend in the sentiment series have almost no impact.\\  
\indent The experiments performed in the contemporaneous (see Appendix \ref{ap:Contemporaneous_quantile} in the supplementary material) and one-lag cases show that the MLSS model is the best model to capture the return variations. For this reason, for the multi-period analysis we will only consider the MLSS model.\\
Considering a general $h$, we wonder if extra lags can add explanatory power to the regression exercise. Using the functional form 
\vspace{-0.4cm}
\begin{equation*}\label{eq:h_lagged_quant_reg}
\hat{V}^{h,\text{MLSS}}\left(\tau\right) =\min_{\left(\alpha^0, \alpha^1 \beta^1\in \mathbb{R}^{6} , \beta^2 \in \mathbb{R}^{6(h-1)}\right)} \sum_{t=h+1}^{T} \rho_{\tau}\left(r^{m}_t - \alpha^0 - \alpha^1 r^{m}_{t-1} - \beta^{1} S_{t-1}^{\text{MLSS}} - \beta^2 \mathcal{L}_{h-1}(S^{\text{MLSS}}_{t-1}) \right)\,,
\vspace{-0.4cm}
\end{equation*}
we separate the contributions given by the first and higher order lags. Under the null hypothesis that $\beta^{2}=0$, the statistic
\vspace{-0.4cm}
\begin{equation}\label{eq:statistic_h_lagged_quant_reg}
L^{h,\text{MLSS}}_{t-h} (\tau) = \frac{2(\hat{V}^{1,\text{MLSS}}(\tau) - \hat{V}^{h,\text{MLSS}}(\tau) )}{\tau(1-\tau) s(\tau) } \to \chi^2_{6(h-1)}\,.
\vspace{-0.4cm}
\end{equation}
Following \citep{Tetlock2007, Garcia2013}, we fix a maximum number of $h=5$ and Table \ref{table:p_val_MLSS_h_lagged} reports the p-values for the different values of $h$.
\begin{table}                                                    
\centering            
{\small                                           
\begin{tabular}{cllll}                                     
\hline                                                           
$\tau$ & $h=2$ & $h=3$ & $h=4$ & $h=5$ \\  
\hline                                                           
$0.01$ & $18.133 \%$ & $29.136 \%$ & $57.652 \%$ & $72.784 \%$ \\                
$0.05$ & $\bf{0.618 \%}$ & $\bf{0.946 \%}$ & $\bf{4.317 \%}$ & $\bf{3.009 \%}$ \\
$0.10$ & $\bf{0.907 \%}$ & $\bf{0.773 \%}$ & $\bf{4.341 \%}$ & $\bf{1.968 \%}$ \\
$0.33$ & $65.530 \%$ & $47.389 \%$ & $74.932 \%$ & $74.071 \%$ \\                
$0.50$ & $62.489 \%$ & $70.078 \%$ & $80.725 \%$ & $90.581 \%$ \\                
$0.66$ & $43.722 \%$ & $53.518 \%$ & $52.853 \%$ & $74.962 \%$ \\                
$0.90$ & $\bf{4.831 \%}$ & $\bf{0.662 \%}$ & $\bf{0.063 \%}$ & $\bf{0.208 \%}$ \\
$0.95$ & $12.800 \%$ & $\bf{2.504 \%}$ & $\bf{0.628 \%}$ & $\bf{2.468 \%}$ \\    
$0.99$ & $38.580 \%$ & $71.448 \%$ & $81.945 \%$ & $87.196 \%$ \\ 
\hline                                                           
\end{tabular}                                                    
\caption{\small p-values for the statistics defined in equation \eqref{eq:statistic_h_lagged_quant_reg} for different values of $h$. Bold values correspond to $\beta^{2}$ significantly different from zero.}                                         
\label{table:p_val_MLSS_h_lagged}  
}                                     
\end{table}                                                      
The $h$-lagged sentiment series are uninformative in the median region, where the one lag sentiment have less explanatory power too. However, in agreement with \citep{Garcia2013}, the lagged sentiment remains informative for few days and, in our case, this is true for the 5\%, 10\%, 90\%, and 95\% quantile levels. It is worth noticing that the 1\% and 99\% quantiles are unaffected by higher-order lags. This shows that, in case of very good or very bad days, the returns are strongly driven by very fresh news ($h=1$) while the older news have no informative power.
\vspace{-0.5cm}

\section{Portfolio allocation with sentiment data}\label{section:portfolio}

This section details an economic application of the MLSS model in portfolio selection and benckmarks the results against a buy-and-hold strategy. We consider the equally weighted portfolio in equation \eqref{eq:def_market_portfolio} and the five filtered signals $S^{MLSS}_t$, $S^{LSS}_t$, $S^{MLNSL}_t$, $S^{LNSL}_t$ and  $S^{Obs}_t$ introduced in the previous section. It is worth noticing that \citep{beckers2018social} and \citep{Garcia2013} showed that the predictability power of the sentiment series declined after 2007. For this reason, we want to challenge the filtering techniques to predict the future daily returns in the time window 2007-2019.

In the first part of this section, we use the sentiment signals as exogenous variables to build a simple classifier and we introduce five trading strategies based on the five sentiment time series. Then, we test these strategies on the February $2007$ - June $2017$ window. This period offers a large series with different economic conditions. The sentiment models are estimated in the same time window. The estimation of multivariate models (MLSS and MLNLS) employs a backward looking technique based on smoothing recursions. Then, one may argue that for the multivariate case the estimation technique may introduce some sort of forward looking bias. We claim that this bias, if any, is negligible and we perform a robustness check where we use the parameter values from February $2007$ - June $2017$ period to filter the TRMI sentiment series from July $2017$ to December $2019$. In this way, the trading signals cannot be affected by any forward looking bias. The results in the out-of-sample period confirm those from February $2007$ - June $2017$, showing that the trading strategies built on the MLSS model are the best performers. The details of the robustness check can be found in Appendix \ref{ap:robustness_check} of the supplementary material.

\subsection{Trading strategies}

In the financial literature, several papers support the strong out-of-sample performance of the equally weighted portfolio (e.g. \citealt{demiguel2009optimal}). The $1/n$ portfolio is used as a baseline for our trading strategies and the long passive position in this portfolio is called \textit{buy-and-hold} strategy. Given that the buy-and-hold portfolio offers a good out-of-sample performance, we assume an investor who only deviates from the baseline strategy if a strong signal which predicts a negative return arrives from the sentiment series. For this reason, the criterion variable needs to capture the behavior of the left tail of returns distribution. We define the criterion binary variable as
\vspace{-0.4cm}
\begin{equation*}
Y_t = \left\{
\begin{array}{lr}
1, & \text{for } \tilde{r}^{m}_t < z_{1/3} \\
0, & \text{otherwise }
\end{array}\right.
\vspace{-0.4cm}
\end{equation*}
where $z_{1/3}$ is the $1/3$ Gaussian quantile and $\tilde{r}^{m}_t = r^{m}_t/\sqrt{RV_t}$ are the standardized market returns with the realized variance, $RV_t$, evaluated by means of $5$-minute intraday returns. The standardization of the returns is crucial to eliminate possible effects due to the persistence of volatility. The choice of the $33\%$ quantile is consistent with the findings of Section \ref{sec:Lagged_quantile}. Moreover, it is a balance between a more conservative choice -- a smaller quantile only sensitive to more extreme and predictive events -- and a larger quantile, which provides a larger number of selling signals but less predictive power. 

\begin{figure}[htbp]
\hspace*{-1.0cm}
\begin{center}
\includegraphics[scale=0.3]{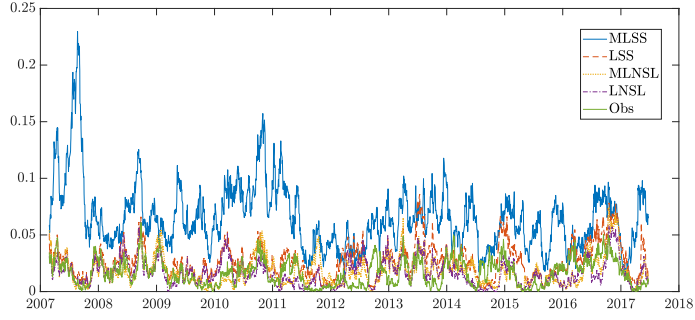} 
\caption{\small McFadden's $R^2$ for the different filtering methods using negative abnormal returns.} \label{fig:R2_negative}
\end{center}
\end{figure}

Since the goal of this paper is to show that the choice of the filtering procedure is essential, a simple classification technique is used. As a classifier, we consider the following conditional logit model
\begin{equation}\label{eq:logit_model}
P\left(Y_{t+1} = 1 | X_t\right) = \text{logit}\left(X^{\text{mod}}_t \theta \right) ,
\end{equation} 
where $\text{logit}(X_t \theta) = \frac{e^{X_t \theta} }{1+e^{X_t \theta} } $ and $X^{\text{mod}}_t = \left[1, \tilde{r}^{m}_t, S^{\text{mod}}_t \right]$. We recall that $S^{\text{mod}}_t$ is a vector whose dimension depends on the filtering model. For further details see the first part of Section \ref{section:applications}.\\
The predicted binary value is defined as
\vspace{-0.3cm}
\begin{equation}\label{eq:s_definition}
\hat{Y}^{\text{mod}}_{t+1} = \left\{
\begin{array}{lr}
1, & \text{for logit} (X^{\text{mod}}_t \theta) > 0.5\\
0, & \text{otherwise }\,.
\end{array}\right.
\vspace{-0.3cm}
\end{equation}
The main advantages of the conditional logit model are twofold. On one hand, the conditional logit model can be easily estimated using MLE. On the other hand, we can easily assess the fitness of the model on the data using the Mc Fadden's $R^2$ measure defined in \citep{mcfadden1973conditional} as
\begin{equation*}
R^2 = 1-\frac{\log(L_m) }{\log(L_0) } \in \left[0, 1\right].
\end{equation*}
$L_m$ represents the maximum likelihood of the complete model \eqref{eq:logit_model} and $L_0$ is the maximum likelihood of the bare model based only on the intercept. The models are estimated using overlapping rolling windows of 6 months (126 observations). We verified that this choice is sufficient to capture the time-varying nature of the explanatory power of the sentiment series. Figure \ref{fig:R2_negative} shows the value of $R^2$ over the February $2007$ - June $2017$ period. The MLSS model has the highest $R^2$ w.r.t the other models, which typically translates in a higher predictive power. In addition, the MLSS $R^2$ has a high variability, suggesting that the predictive power changes through time. This latter finding suggests that the sentiment signal can be a good returns predictor in certain periods and a poor predictor in others. This intuition will be exploited later to generate trading strategies based on the $R^2$.
The estimated $\bar{Y}_t^{\text{mod}}$ defined in \eqref{eq:s_definition} translates in the trading signal
\vspace{-0.4cm}
\begin{equation}\label{eq:trading_signal_short_selling}
s^{\text{mod}}_{t+1} = \left\{
\begin{array}{lr}
1, & \text{if } \hat{Y}_{t+1}^{\text{mod}} = 0\\
-1, & \text{if } \hat{Y}_{t+1}^{\text{mod}} = 1
\end{array}\right.
\vspace{-0.4cm}
\end{equation}
where $s^{\text{mod}}_{t+1} = 1$ ($s^{\text{mod}}_{t+1} = -1$) represents a buy (sell) signal in the equally weighted portfolio \eqref{eq:def_market_portfolio}. At any day $t$, at the closing time of the trading day, the investor uses the sentiment signal $S^{\text{mod}}_t$ and the standardized realized daily returns $\tilde{r}^{m}_t$ to forecast the binary variable $\hat{Y}_{t+1}^{\text{mod}}$ and the relative trading signal. Naming $c_0$ the number of shares bought or sold in any transaction, there are three possible scenarios
\begin{enumerate}
\item{$s^{\text{mod}}_{t} = s^{\text{mod}}_{t+1}$ : In this case the prediction on the future realization does not change and the investor does not re-balance the portfolio.}
\item{$s^{\text{mod}}_{t} = +1$ and $s^{\text{mod}}_{t+1} = -1$ : The investor had a long position in the equally weighted portfolio at time $t$ but the prediction changed. She sells the current position and short sells $c_0$ shares of the same portfolio.}
\item{$s^{\text{mod}}_{t} = -1$ and $s^{\text{mod}}_{t+1} = +1$ : The investor had a short position in the equally weighted portfolio at time $t$ but the prediction changed. She buys $2c_0$ shares of the portfolio.}
\end{enumerate}
Please notice that the only exception is for $s^{\text{mod}}_{1}$ because we initialized $s^{\text{mod}}_{0} = 0$. In this case, the equally weighted portfolio is bought when $s^{\text{mod}}_{1} = 1$ and it is short sold when $s^{\text{mod}}_{1} = -1$. The investor's portfolio is then built as
\vspace{-0.3cm}
\begin{equation} \label{eq:definition_portfolio_costs}
\begin{cases}
P^{\text{mod}}_{t+1} = s_{t+1}^{\text{mod}}c_0 M_{t+1} + \text{cash}_{t+1},\\
\text{cash}_{t+1} = \text{cash}_{t} - (s_{t+1}^{\text{mod}}-s_{t}^{\text{mod}})c_0 M_{t+1} - |s_{t+1}^{\text{mod}}-s_{t}^{\text{mod}}|c_0 M_{t+1} \frac{\text{cost}}{2}\,,
\end{cases}
\vspace{-0.3cm}
\end{equation} 
where $cost$ is the percentage trading cost and $M_t$ is defined in \eqref{eq:def_market_portfolio}. The first equation in \eqref{eq:definition_portfolio_costs} shows that the value of the portfolio is composed by the value of the invested amount $s_{t+1}^{\text{mod}}c_0 M_{t+1}$ plus the cash position. The latter increases when $s_{t+1}^{\text{mod}}<s_{t}^{\text{mod}}$, meaning that the investor sells the portfolio and receives cash, and decreases when $s_{t+1}^{\text{mod}}>s_{t}^{\text{mod}}$, meaning that the investor buys and erodes the cash position. The second equation includes the impact of the transaction costs. Specifically, every time that a transaction happens, i.e. $s_{t+1}^{\text{mod}}\neq s_{t}^{\text{mod}}$, the investor pays an extra cost proportional to the current value of the equally weighted portfolio $M_{t+1}$.

We fix the starting point  $s_{0}^{\text{mod}} = 0$, $\text{cash}_{0} = 100,000\$$ and the parameter $c_0 = 100,000\$/M_0$. In the paper we only report the results for the case with trading costs, while the results with zero trading costs are reported in Appendix \ref{ap:without_trading_costs} of the supplementary material. From now on, we refer to \textit{without trading costs} when the portfolio in equation (\ref{eq:definition_portfolio_costs}) is evaluated with cost $ = 0$ and to \textit{with trading costs} when costs $ = 0.1\%$ as in \citep{gilli2009empirical} and \citep{avellaneda2010statistical}. In the following sections, the number of transactions is evaluated as $Tr^{\text{mod}} = \sum_{i=0}^{T-1} |s_{i+1}^{\text{mod}} - s_{i}^{\text{mod}}|$
and the transaction costs are evaluated as $Tc^{\text{mod}} = \sum_{i=0}^{T-1} |s_{i+1}^{\text{mod}} - s_{i}^{\text{mod}}|c_0 M_{i+1} \frac{\text{cost}}{2}$.
It is worth noticing that the change of signal effectively produces two transactions. For instance, if the signal moves from $s_t = 1$ to $s_{t+1} = -1$, the first transaction is the liquidation of the long position and the second transaction is the short position on the asset. In addition, most of the time the selling signal appears for only one day and disappears the day after. Then, the typical path of a selling signal is given by $s_t = 1$, $s_{t+1} = -1$ and $s_{t+2} = 1$ producing a total of four transactions.

The transaction costs can strongly depress the overall performance of the portfolio. To partially mitigate this drawback, we can decrease the number of transactions using the McFadden's $R^2$ as a measure of the reliability of the signal $\hat{Y}^{\text{mod}}_{t}$. We compute the empirical quantile $z_{\alpha}^{1,t}\left(R^2\right)$ of the McFadden $R^2$ over the time window $(1,\cdots,t)$. The quantile $z_{\alpha}^{1,t}\left(R^2\right)$ is $\mathcal{F}_t$-measurable and does not introduce a forward looking bias. We can reduce the number of trades conditioning the selling signal at time $t$ on the level of the McFadden's $R^2$ evaluated in the previous 6 months. The $R^2$ adjusted trading signal is then defined as follows
\vspace{-0.4cm}
\begin{equation}\label{eq:R_squared_correction2}
\bar{s}^{\text{mod}}_t = \left\{
\begin{array}{lr}
-1, & \text{if }\hat{Y}_{t+1}^{\text{mod}} = 1 \text{ and } R^{2,\text{mod}}_{t} \geqslant z_{\alpha}^{1,t}\left(R^{2,\text{mod}}\right) \\
1, & \text{otherwise }\,.
\end{array}\right.
\vspace{-0.4cm}
\end{equation}
The value $\alpha$ determines the reduction in the number of trades. The higher $\alpha$ is, the smaller is the number of transactions. The five strategies, together with the buy-and-hold strategy itself, are evaluated according to six measures, the \textit{annual return}, the \textit{annual volatility}, the \textit{annual negative volatility}, the \textit{Sharpe ratio}, the \textit{Sortino ratio}, and the \textit{maximum drawdown} (MDD). In the next section, in a first step, the portfolios with the trading signals \eqref{eq:trading_signal_short_selling} with and without trading cost are analysed. Then, we assess the impact and the performance of the trade reduction strategy based on \eqref{eq:R_squared_correction2}.

\subsection{Empirical application: February 2007 - June 2017}\label{sec:2007_2017}
The 2007-2009 crisis and the 2009-2017 bull market are good backtesting periods for the sentiment portfolios because we can test the return predictability during different market conditions.

\indent Table \ref{table:portfolio_market_short_selling_cost} reports the performances of the five sentiment strategies together with the buy-and-hold portfolio with trading costs. The sentiment-based strategies have, excluding the LNSL and the Obs, a smaller volatility and MDD than the buy-and-hold portfolio. In addition, the MLSS portfolio produces returns similar to the buy-and-hold strategy, lower negative volatility, and consequently higher Sharpe and Sortino ratios than all the other strategies. The lower performance for the annual returns is due to the higher transaction costs. Indeed in Appendix \ref{ap:without_trading_costs} of the supplementary material we show that, when the trading costs are not considered, the MLSS strategy produces higher annual returns than all the other strategies. In addition, when we compare \textit{without trading costs} experiment with the \textit{with trading costs} experiment, the excessive number of transactions for the MLSS strategy  reduces the Sharpe ratio gain with respect to the buy-and-hold portfolio from $40\%$ to $10\%$ and the Sortino ratio gain from $48\%$ to $16\%$. 
In Appendix \ref{ap:portfolio_significance} of the supplementary material we show that the selling signal generated by the MLSS sentiment series corresponds to statistically significant returns predictability.
\begin{table}
\centering
{\small
\begin{tabular}{c|llllll}
\hline
Measures & BH & MLSS & LSS & MLNSL & LNSL & Obs \\
\hline
A. return (\%) & \textbf{8.975} & $7.891 $ & $6.977 $ & $8.882 $ & $8.143 $ & $6.986 $ \\
A. volatility (\%) & $19.132 $ & \textbf{15.209} & $18.136 $ & $17.952 $ & $19.431 $ & $19.955 $ \\         
A. neg. volatility (\%) & $15.523 $ & \textbf{11.767} & $14.339 $ & $14.474 $ & $15.374 $ & $16.055 $ \\
A. Sharpe ratio & $0.469 $ & \textbf{0.519} & $0.385 $ & $0.495 $ & $0.419 $ & $0.35 $ \\                   
A. Sortino ratio & $0.578 $ & \textbf{0.671} & $0.487 $ & $0.614 $ & $0.53 $ & $0.435 $ \\ 
MDD (\$) & $59377 $ & $57235 $ & $50335 $ & \textbf{49773} & $63595 $ & $62785 $ \\                    
Number of trades & $1 $ & $553 $ & $161 $ & $81 $ & $73 $ & $93 $ \\
Transaction costs (\$) & $50 $ & $37974 $ & $14866 $ & $5565 $ & $4085 $ & $7544 $ \\  
\hline
\end{tabular}
\caption{\small  Performances of the six strategies with transaction cost for the period February $2007$ - June $2017$. In bold, the best performance per row. BH is the buy-and-hold portfolio, while MLSS, LSS, MLNSL, LSNSL, and Obs correspond to portfolios built from the corresponding model for the sentiment time series.}
\label{table:portfolio_market_short_selling_cost}
}
\end{table}
The transaction costs incurred by the MLSS portfolio throughout the nine years amount in total to $38\%$ of the starting capital. For this reason, we employ the trading signal $\bar{s}^{\text{MLSS}}$ defined in equation \eqref{eq:R_squared_correction2}, which penalizes signals with moderate McFadden's $R^2$. Table \ref{table:portfolio_market_short_selling_R2} reports the performances of the strategies based on the penalized signal for different values of $\alpha$. As expected, the higher the value of $\alpha$ and the lower the number of transactions is. In addition, the $R^2$-based signal produces higher quality signal and effectively increases the performance of the portfolios. The number of transactions decreases almost linearly but the Sharpe and Sortino ratios strongly increase. They reach a maximum value when $\alpha = 0.65$. These findings further corroborate the intuition that the MLSS sentiment strongly anticipates future returns during the financial crisis, given that the $R^2$ values in figure \ref{fig:R2_negative} are higher than the unconditional average during the $2007-2009$ period. Again this feature is peculiar for the MLSS filter while no evidence of return predictability is reported for the other filtering techniques. Again, the statistical significance of these strategies is reported in Appendix \ref{ap:portfolio_significance} of the supplementary material.

\begin{table}                                                                  \centering                                                                     \small{
\begin{tabular}{c|lllllll}                                                     \hline                                                                         Measures & BH & $\alpha = 0\%$ & $\alpha = 20\%$ & $\alpha = 35\%$ & $\alpha = 50\%$ & $\alpha = 65\%$ & $\alpha = 80\%$ \\
\hline                                                                         A. return (\%) & $8.975 $ & $7.891 $ & $8.225 $ & $9.575 $ & $9.84 $ & \textbf{10.248} & $9.184 $ \\                
A. volatility (\%) & $19.132 $ & $15.209 $ & $15.679 $ & $14.083 $ & $13.601 $ & \textbf{13.538} & $17.201 $ \\      
A. neg. volatility (\%) & $13.601 $ & $10.888 $ & $11.196 $ & $9.901 $ & $9.511 $ & \textbf{9.443} & $12.216 $ \\
A. Sharpe ratio & $0.469 $ & $0.519 $ & $0.525 $ & $0.680 $ & $0.723 $ & \textbf{0.757} & $0.534 $ \\                 
A. Sortino ratio & $0.660 $ & $0.725 $ & $0.735 $ & $0.967 $ & $1.035 $ & \textbf{1.085} & $0.752 $ \\                
MDD (\$) & $59377 $ & $57235 $ & $63522 $ & $49160 $ & \textbf{33264} & $35600$ & $59486 $ \\                  
Number of trades & $1 $ & $553 $ & $437 $ & $349 $ & $273 $ & $169 $ & $57 $ \\ Transaction costs (\$) & $50 $ & $37974 $ & $30626 $ & $25283 $ & $20074 $ & $12007 $ & $4127 $ \\      
\hline 
\end{tabular}                                                                
\caption{\small Performances of the MLSS based strategies built from equation \eqref{eq:R_squared_correction2} for different values of $\alpha$  $\times 100$. BH is the buy-and-hold portfolio. In bold, the best performance per row.}                                                          \label{table:portfolio_market_short_selling_R2}
}
\end{table}
\vspace{-0.5cm}

\section{Conclusions}\label{section:conclusion}

In this paper, we presented a novel way to filter multivariate sentiment time series. The approach is very general and encompasses previous models discussed in the literature. Using a dynamic factor model, we were able to identify two different sentiment components. The first one, named long-term sentiment and modeled as a random walk, captures the common trends which drive the long-term dynamics. The second component, dubbed short-term sentiment and modeled as a VAR(1) process, captures short-term swings of market mood.  
An extensive empirical section investigates the different features of the two sentiment components. In a first analysis, we pointed out that one of the long-term sentiment factors co-integrates with the first principal component of the market. Quite surprisingly, the structure of the sentiment factor loadings does not mimic the typical uniform profile of the market factor. Some assets are over-expressed and contribute to the factor with a positive or negative sign, while others are under-expressed. Concerning the short-term sentiment, its multivariate dependence structure explains a sizable fraction of the residual covariance in a single factor market model. This result suggests that the short-term component captures transient and rapidly changing trends associated with the idiosyncratic components of the market. 
In a second analysis, based on quantile regression, we showed that the Multivariate Long-Short Sentiment model provides the highest explanatory power of lagged and contemporaneous returns. Essential to achieve statistical significance are the multivariate nature of the approach and the separation of the sentiment signal in a long and a short component. In particular, disentangling the short-term sentiment is crucial to capture the behavior of extreme returns. In a further analysis, we observed that newspapers and social media differently react to negative and positive returns. Specifically, they can effectively explain abnormal returns from one to five days in advance, but they almost immediately digest the positive market realizations while they echo negative realizations for several days to come.\\
It is worth noting that \citep{Tetlock2007} and \citep{Garcia2013} reported results similar to ours for the unfiltered sentiment focusing on period before 2007. Using the TRMI dataset, \citep{beckers2018social} showed that the forecasting power on returns of the sentiment dropped dramatically after 2007. Our results suggest that the filtering procedures are more important nowdays than in the past. Consistently, in a final investigation, we performed an asset allocation exercise where the selling signal are based on the sentiment series. In line with results from the quantile regression, the portfolio based on the MLSS filter significantly outperforms the benchmark buy-and-hold strategy and the other strategies based on different filtering techniques. 
\vspace{-0.5cm}

\section*{Supplementary materials}
{\small
The supplementary materials include the details of the estimation procedure in Appendix \ref{ap:estimation} as well as the details of Kalman filter and smoother in Appendix \ref{ap:filter} and the equations of the Expectation Maximization algorithm in Appendix \ref{ap:expectation_max}. Appendix \ref{ap:tickers_description} provides an overview of the stocks used in the empirical analysis. Appendix \ref{ap:snr} compares the different signal-to-noise ratios filtered by the MLSS and MLNSL model. Appendix \ref{ap:Contemporaneous_quantile} investigates the contemporaneous relation between the sentiment and return series. Appendices \ref{ap:without_trading_costs}, \ref{ap:robustness_check}, and \ref{ap:portfolio_significance} report the \textit{without trading costs} analysis, the robustness check, and the statistical significance of the portfolio allocation exercise, respectively.
}
\vspace{-0.5cm}

\section*{Acknowledgments}
{\small
The authors thank Thomson Reuters for kindly providing Thomson Reuters MarketPsych Indices time series. We benefited from discussion with Giuseppe Buccheri, Fulvio Corsi, Luca Trapin, as well as with conference participants to the Quantitative Finance Workshop 2019 at ETH in Zurich and the AMASES XLIII Conference in Perugia.
}

{\small
\onehalfspacing
\bibliography{biblio}
\bibliographystyle{elsarticle-harv}
}

\newpage
{\huge \textbf{Supplementary Material}}
\appendix
\section{Estimation procedure}\label{ap:estimation}
The estimation of model \eqref{eq:sentiment_model} is performed using the Kalman filter \citep{KalmanFilter} and the Expectation Maximization (EM) method in \citep{dempster1977maximum} and \citep{Shumway1982} which was proposed to deal with incomplete or latent data and intractable likelihood. The EM algorithm is a two-step estimator. In the first step, we write the likelihood considering the latent process as observed. In the second step, we re-estimate the static parameters maximizing the expectation obtained in the first step. This routine is repeated until some convergence criterion is satisfied.\\
\indent To cast \eqref{eq:sentiment_model} in a standard state-space representation, we use the same procedure of \citep{Banbura} and define the augmented states $\tilde{\Lambda}$, $\tilde{F}$, $\tilde{\Phi}$ and $\tilde{Q}$ s.t.
\begin{equation}\label{eq:sentiment_model_augmented}
\begin{aligned}
S_t &= \tilde{\Lambda} \tilde{F}_t  + \epsilon_t, \qquad \epsilon_t \sim \mathcal{N}\left(0,R \right),\\
\tilde{F}_{t} &= \tilde{\Phi} \tilde{F}_{t-1} + v_t , \quad v_t \sim \mathcal{N}\left(0, \tilde{Q} \right),
\end{aligned}
\end{equation}
where
\begin{equation}\label{eq:DFA_tilde_lambda}
\tilde{\Lambda} = \begin{bmatrix}
    \Lambda & I_{K}
\end{bmatrix} \in \mathbb{R}^{K\times \left(q+K \right) }
\end{equation}
\begin{equation}\label{eq:DFA_tilde_F}
\tilde{F}_t = \begin{bmatrix}
    F_t\\
    \Psi_t
\end{bmatrix} \in \mathbb{R}^{\left(q+K \right)\times 1 }
\end{equation}
\begin{equation}\label{eq:DFA_A}
\tilde{\Phi} = \begin{bmatrix}
    I_q & 0\\
    0 & \Phi
	\end{bmatrix} \in \mathbb{R}^{\left(q+K \right)\times \left(q+K \right) }
\end{equation}
\begin{equation}\label{eq:DFA_tilde_Q}
\tilde{Q} = \begin{bmatrix}
    Q_{long} & 0\\
    0 & Q_{short}
\end{bmatrix} \in \mathbb{R}^{\left(q+K \right)\times \left(q+K \right)}
\end{equation}
The EM renders the approach feasible in high dimension. Indeed, while a direct numerical maximization of the likelihood is computationally demanding, the EM algorithm, thanks to the Kalman filtering and smoothing recursions, can be formulated in closed-form. See Appendix \ref{ap:filter} and \ref{ap:expectation_max}. In particular, it allows to disentangle the long-term sentiment $F_t$ and the short-term sentiment $\Psi_t$. To derive the EM steps we consider the log-likelihood $l\left(S_t,\tilde{F}_t,\theta\right)$ where $\theta$ denotes the set of static parameters $\tilde{\Lambda}$, $\tilde{\Phi}$, $\tilde{Q}$ and $R$. The EM proceeds in a sequence of steps:
\begin{enumerate}
\item{E-step: it evaluates the expectation of the log-likelihood using the estimated parameters from the previous iteration $\theta\left(j \right)$:
\[G\left(\tilde{\Lambda}\left(j\right), \tilde{\Phi}\left(j\right),\tilde{Q}\left(j\right), R\left(j\right)\right) = E\left[l\left(S_t,\tilde{F}_t,\theta\left(j\right) \right)| S_1,\ldots,S_T \right]. \]
The E-step strongly relies on the Kalman smoother. The details are explained in Appendix \ref{ap:filter}.
}
\item{M-step: the parameters are estimated again maximizing the expected log-likelihood with respect to $\theta$:
\[\theta\left(j+1 \right) = \text{arg}\max_{\theta} G\left(\tilde{\Lambda}\left(j\right), \tilde{\Phi}\left(j\right),\tilde{Q}\left(j\right), R\left(j\right)\right). \]
The M-step is performed updating the static parameters. Further information on the equations can be found in Appendix \ref{ap:expectation_max}.
}
\end{enumerate}
We initialize the parameters $\theta\left(0\right)$ and repeat steps 1 and 2 until we reach the convergence criterion 
\begin{equation}\label{eq:convergence_criteria}
\frac{|l\left(S_t,\tilde{F}_t,\theta\left(j\right) \right) - l\left(S_t,\tilde{F}_t,\theta\left(j-1\right) \right)|}{|l\left(S_t,\tilde{F}_t,\theta\left(j\right) \right) + l\left(S_t,\tilde{F}_t,\theta\left(j-1\right) \right)|}<\frac{\epsilon}{2}.
\end{equation}
We set $\epsilon = 10^{-3}$.\\
\indent As observed in \citep{harvey_1990}, the dynamic factor model \eqref{eq:sentiment_model_augmented} is not identifiable. Indeed, if we consider a non singular invertible matrix $M$, then the parameters $\theta_1 = \{\Lambda, R, Q \}$ and $\theta_2 = \{\Lambda M^{-1}, R, MQM' \}$ are observationally equivalent, then starting from $S_t$ we cannot distinguish $\theta_1$ from $\theta_2$. We solve this identification problem using the approach proposed by \citep{harvey_1990}, imposing the following restrictions
\begin{equation}\label{eq:identification_harvey}
\begin{aligned}
\tilde{Q} &= \begin{bmatrix}
    I_q & 0\\
    0 & Q_{short}
\end{bmatrix} \\
\Lambda &= \begin{bmatrix}
    \lambda_{11} & 0 & 0 & \dots  & 0 \\
    \lambda_{21} & \lambda_{22} & 0 & \dots  & 0 \\
    \vdots & \vdots & \vdots & \ddots & \vdots \\
    \vdots & \vdots & \vdots & \vdots & 0 \\
    \vdots & \vdots & \vdots & \vdots & \vdots \\
    \lambda_{K1} & \lambda_{K2} & \lambda_{K3} & \dots  & \lambda_{Kq}
\end{bmatrix}
\end{aligned}
\end{equation}
where $\Lambda$ is the $K\times q$ sub-matrix in \eqref{eq:DFA_tilde_lambda}.\\
\indent The specifications of $\tilde{\Lambda}$, $\tilde{\Phi}$, $\tilde{Q}$ and $R$ in \eqref{eq:DFA_tilde_lambda}, \eqref{eq:DFA_A} and  \eqref{eq:DFA_tilde_Q}, together with the identification restrictions defined in \eqref{eq:identification_harvey}, impose several constraints to the estimations. The EM procedure allows us to impose restrictions on the parameters in a closed-form. 
According to \citep{WU199699} and \citep{Bork2009}, we get the constrained $\tilde{\Phi}$, $\tilde{\Lambda}$, $\tilde{Q}$ and $R$ as:
\begin{equation}\label{eq:Phi_constrained}
\textsf{vec} (\tilde{\Phi}_r) = \textsf{vec}(\tilde{\Phi}) + \left(A^{-1} \otimes \tilde{Q}\right) M (M(A^{-1}\otimes \tilde{Q})M')^{-1}(k_{\Phi}- M \textsf{vec}(\Phi))
\end{equation}
where $A$ is defined in equation \ref{ap:eq:ABCE1E2E3}, $M$ is the $f \times 2K(r+K)$ matrix, $f$ is the number of constraints, $k_{\Phi}$ is the $f$ vector containing the constraints values such that $M \textsf{vec}(\tilde{\Phi}) = k_{\Phi}.$\\
Equivalently, for the restricted $\Lambda_r$:
\begin{equation}\label{eq:Lambda_identific}
\textsf{vec} (\Lambda_r) = \textsf{vec} (\Lambda) + \left(E_1^{-1} \otimes R\right) G (G(E_1^{-1}\otimes R)G')^{-1}(k_{\lambda}- G \textsf{vec}(\Lambda))
\end{equation}
where $E_1$ is defined in \ref{ap:eq:ABCE1E2E3}, $G$ is the $s \times Kr$ matrix, $s$ is the number of constraints, $k_{\lambda}$ is the $s$ vector containing the constraints values such that $G \textsf{vec}(\Lambda) = k_{\lambda}$. The details for the evaluation of $\tilde{Q}$ and $R$ are reported in equation \eqref{ap:eq:EM_update_Q_autocorr} and \eqref{ap:eq:EM_update_R_autocorr} and the restrictions, according to \citep{WU199699}, can be imposed elementwise.\\

The final estimation scheme reads as follows:
\begin{enumerate}
\item{Initialize $\tilde{\Lambda}\left(0\right)$, $\tilde{\Phi}\left(0\right)$, $\tilde{Q}\left(0\right)$ and $R\left(0\right)$
}
\item{Perform the E-step using the estimations $\tilde{\Lambda}\left(j\right)$, $\tilde{\Phi}\left(j\right)$, $\tilde{Q}\left(j\right)$, $R\left(j\right)$ and the Kalman smoother.
}
\item{Perform the M-step and evaluate the new estimators $\tilde{\Lambda}\left(j+1\right)$, $\tilde{\Phi}\left(j+1\right)$, $\tilde{Q}\left(j+1\right)$ and $R\left(j+1\right)$.
}
\item{
Use the unrestricted estimations and \eqref{eq:Phi_constrained} and \eqref{eq:Lambda_identific} to obtain the restricted ones.
}
\item{
Repeat 2, 3 and 4 above until the estimates and the log-likelihood reach convergence.
}
\end{enumerate}
Finally, since the number of long-term sentiment $q$ is considered as known, we select the optimal $q$ using the AIC and BIC indicators.

\newpage

\section{Filter and Smoother recursions}\label{ap:filter}
In this section, we report Kalman Filter and Smoother recursions ancillary to the EM algorithm. The derivation of the formulas which follow can be found in (Shumway and Stoffer, 1982). \\
Starting from system \eqref{eq:sentiment_model_augmented}, we calculate recursively the Kalman Filter as:
\begin{equation}\label{ap:eq:KF}
\begin{aligned}
\tilde{F}_{t|t-1} &= E\left[\tilde{F}_t|S_1,\ldots,S_{t-1} \right] = \tilde{\Phi}\tilde{F}_{t-1|t-1}\\
P_{t|t-1} &= E\left[\left(\tilde{F}_t-\tilde{F}_{t|t-1} \right)\left(\tilde{F}_t-\tilde{F}_{t|t-1} \right)'|S_1,\ldots,S_{t-1} \right] = \tilde{\Phi}P_{t-1|t-1}\tilde{\Phi}'+Q\\
K_t &= P_{t|t-1}\tilde{\Lambda}'\left(\tilde{\Lambda} P_{t|t-1} \tilde{\Lambda}'+R \right)^{-1}\\
\tilde{F}_{t|t} &= \tilde{F}_{t|t-1}+K_t\left(S_t-\tilde{\Lambda}\tilde{F}_{t|t-1} \right)\\
P_{t|t} &= P_{t|t-1}-K_t\tilde{\Lambda}P_{t|t-1}
\end{aligned}
\end{equation}
where we take $\tilde{F}_{0|0} = \mu$ and $P_{0|0} = \Sigma$. Now, using backward recursions $t=T,\ldots,1$ we derive the Smoother as
\begin{equation}\label{ap:eq:KS}
\begin{aligned}
J_{t-1} & =P_{t-1|t-1}\tilde{\Phi}'\left(P_{t|t-1} \right)^{-1}\\
\tilde{F}_{t-1|T} &= \tilde{F}_{t-1|t-1}  + J_{t-1}\left(\tilde{F}_{t|T} - \tilde{\Phi}\tilde{F}_{t-1|t-1} \right)\\
P_{t-1|T} &= P_{t-1|t-1} +J_{t-1}\left(P_{t|T}-P_{t|t-1} \right)J_{t-1}' \\
P_{t-1,t-2|T} &= P_{t-1|t-1}J_{t-2}'+J_{t-1}\left(P_{t,t-1|T}-\tilde{\Phi}P_{t-1|t-1} \right)J_{t-2}'
\end{aligned}
\end{equation}
where $P_{T,T-1|T} = \left(I-K_T\tilde{\Lambda} \right)\tilde{\Phi}P_{T-1|T-1}$.

\newpage

\section{Expectation Maximization}\label{ap:expectation_max}
The log-likelihood of the model \eqref{eq:sentiment_model_augmented} is
\[
\begin{split}
l\left(S_t,\tilde{F}_t,\theta\left(j\right) \right)=& \log f(\tilde{F}_0) + \sum_{t=1}^{T} \log f(\tilde{F}_t|S_{t-1}) + \sum_{t=1}^{T} \log f(S_t|\tilde{F}_t)\\
=& -\frac{1}{2}\log|\Sigma| -\frac{1}{2} \left(\tilde{F}_0 - a\right)\Sigma^{-1}\left(\tilde{F}_0 - a\right)'\\
& -\frac{T}{2}\log|\tilde{Q}|-\frac{1}{2}\sum_{t=1}^{T} \left(\tilde{F}_t - \tilde{\Phi}\tilde{F}_{t-1}\right)\tilde{Q}^{-1} \left(\tilde{F}_t - \tilde{\Phi}\tilde{F}_{t-1}\right)'\\
& -\frac{T}{2}\log|R| -\frac{1}{2}\sum_{t=1}^{T} \left(S_t - \tilde{\Lambda} \tilde{F}_{t}\right)R^{-1} \left(S_t - \tilde{\Lambda} \tilde{F}_{t}\right)'\\
\end{split}
\]
where $a$ and $\Sigma$ are the parameters s.t. $\tilde{F}_0 \sim \mathcal{N}\left(a, \Sigma \right)$.
\subsubsection*{E-step}\label{ap:E_step}
The objective function to maximize is, from (Shumway and Stoffer, 1982),
\[G\left(a,\Sigma,R,\tilde{Q}, \tilde{\Lambda},\tilde{\Phi} \right) = E_{m} \left[\log f|S_1,\ldots,S_T \right],
\]
where $E_m$ denotes the conditional expectation relative to a density containing
the $m$th iterate values $a(m),\Sigma(m),R(m),\tilde{Q}(m), \tilde{\Lambda}(m)$ and $\tilde{\Phi}(m)$.\\
Using now the Kalman  smoother \eqref{ap:eq:KS} we can derive
\[
E\left[\left(S_t - \tilde{\Lambda} \tilde{F}_t\right)\left(S_t - \tilde{\Lambda} \tilde{F}_t\right)'|S_1,\ldots S_T \right] = \left(S_t - \tilde{\Lambda} \tilde{F}_{t|T}\right)\left(S_t - \tilde{\Lambda} \tilde{F}_{t|T}\right)' + \tilde{\Lambda} P_{t|T} \tilde{\Lambda'}
\]
\[
\begin{aligned}
E\left[\left(\tilde{F}_t - \tilde{\Phi}\tilde{F}_{t-1}\right)\left(\tilde{F}_t - \tilde{\Phi} \tilde{F}_{t-1}\right)'|S_1,\ldots, S_T \right] =& P_{t|T} + \tilde{F}_{t|T} \tilde{F}_{t|T}' +\tilde{\Phi}P_{t-1|T}\tilde{\Phi}'\\
&+ \tilde{\Phi}\tilde{F}_{t-1|T}\tilde{F}_{t-1|T}'\tilde{\Phi}'-P_{t,t-1|T}\tilde{\Phi}'\\
&-\tilde{F}_{t|T}\tilde{F}_{t-1|T}'\tilde{\Phi}'
-\tilde{\Phi}P_{t,t-1|T}-\tilde{\Phi}\tilde{F}_{t-1|T}\tilde{F}_{t|T}',
\end{aligned}
\]
lead to
\begin{equation}\label{G_DFA}
\begin{aligned}
G\left(a,\Sigma,R,\tilde{Q}, \tilde{\Lambda},\tilde{\Phi} \right) = & -\frac{1}{2}\log|\Sigma| -\frac{1}{2} tr\{\Sigma^{-1}\left[P_{0|T}+\left(\tilde{F}_0 - a\right)\left(\tilde{F}_0 - a\right)'\right] \}\\
&-\frac{T}{2}\log|\tilde{Q}|-\frac{1}{2}tr\{\tilde{Q}^{-1} \left(C-B\tilde{\Phi}' - \tilde{\Phi}B' +\tilde{\Phi}A\tilde{\Phi}' \right)\}\\
&-\frac{T}{2}\log|R|-\frac{1}{2}tr\{R^{-1} \left(E_3-\tilde{\Lambda} E_2' - E_2\tilde{\Lambda}' + \tilde{\Lambda} E_1 \tilde{\Lambda}' \right)\},
\end{aligned}
\end{equation}
where
\begin{equation}\label{ap:eq:ABCE1E2E3}
\begin{aligned}
A &= \sum_{t=1}^{T}\left(\tilde{F}_{t-1|T}\tilde{F}_{t-1|T}' +P_{t-1|T} \right),\\
B &= \sum_{t=1}^{T}\left(\tilde{F}_{t|T}\tilde{F}_{t-1|T}' +P_{t,t-1|T} \right),\\
C &= \sum_{t=1}^{T}\left(\tilde{F}_{t|T}\tilde{F}_{t|T}' +P_{t|T} \right),\\
E_1 &= \sum_{t=1}^{T}P_{t|T} + \tilde{F}_{t|T}\tilde{F}_{t|T}',\\
E_2 &= \sum_{t=1}^{T} S_{t} \tilde{F}_{t|T}',\\
E_3 &= \sum_{t=1}^{T}S_{t}S_{t}'.
\end{aligned}
\end{equation}
\subsubsection*{M-step}\label{ap:M_step}
The resulting update equations are
\begin{equation}\label{ap:eq:EM_update_Lambda_autocorr}
\Lambda(m+1) = E_2 E_1^{-1}
\end{equation}
\begin{equation}\label{ap:eq:EM_update_Phi_autocorr}
\tilde{\Phi}(m+1) = B A^{-1}
\end{equation}
\begin{equation}\label{ap:eq:EM_update_Q_autocorr}
\tilde{Q}(m+1) = \frac{1}{T}\left(C-B\tilde{\Phi}\left(m+1\right)'-\tilde{\Phi}\left(m+1\right)B'+\tilde{\Phi}\left(m+1\right)A\tilde{\Phi}\left(m+1\right)'\right)
\end{equation}
\begin{equation}\label{ap:eq:EM_update_R_autocorr}
R(m+1) = \frac{1}{T}\left(E_3-\tilde{\Lambda}(m+1)E_2'-E_2\tilde{\Lambda}(m+1)'+\tilde{\Lambda}(m+1)E_1\tilde{\Lambda}(m+1)' \right)
\end{equation}
\begin{equation}\label{ap:eq:EM_update_a_autocorr}
a(m+1) = \tilde{F}_{0|T}
\end{equation}
\begin{equation}\label{ap:eq:EM_update_Sigma_autocorr}
\Sigma(m+1) = P_{0|T}\,.
\end{equation}
For simplicity, in our estimations we impose $\tilde{F}_0 = 0$.

\newpage

\section{List of stocks}\label{ap:tickers_description}
Table \ref{table:tickers_sectors} reports names and sectors of the $27$ stocks considered in the empirical analysis. 
\begin{table}[!htbp]                                                                                                                                                                                                                                                           
\centering     
{\small                                    
\begin{tabular}{cccc}                         
\hline                                             
Tickers & Name & Sector ticker & Sector name \\     
\hline                                             
VZ & Verizon & COM & Communication Services \\     
CVX & Chevron & ENE & Energy \\                    
AXP & American Express Company & FIN & Financial \\
GS & Goldman Sachs & FIN & Financial \\            
JPM & JPMorgan Chase & FIN & Financial \\          
JNJ & Johnson \& Johnson & HLC & Health Care \\     
MRK & Merck & HLC & Health Care \\                 
PFE & Pfizer & HLC & Health Care \\                
UNH & UnitedHealth & HLC & Health Care \\          
BA & Boeing & IND & Industrials \\                 
CAT & Caterpillar & IND & Industrials \\           
GE & General Electric & IND & Industrials \\       
MMM & 3M Co & IND & Industrials \\                 
UTX & United Technologies & IND & Industrials \\   
XOM & XOMA Corp & MAT & Basic Materials \\         
KO & Coca-Cola & NCY & Consumer Goods \\           
PG & Procter \& Gamble & NCY & Consumer Goods \\    
AAPL & Apple & TEC & Technology \\                 
CSCO & Cisco & TEC & Technology \\                 
IBM & IBM & TEC & Technology \\                    
INTC & Intel & TEC & Technology \\                 
MSFT & Microsoft & TEC & Technology \\             
DIS & Disney & YCY & Consumer Cyclical \\          
HD & Home Depot & YCY & Consumer Cyclical \\       
MCD & McDonalds & YCY & Consumer Cyclical \\       
NKE & Nike & YCY & Consumer Cyclical \\            
WMT & Wal-Mart & YCY & Consumer Cyclical \\        
\hline                                             
\end{tabular}                                     
\caption{List of investigated stocks, their ticker, and the economic sector according to the classification of Yahoo Finance.}                           
\label{table:tickers_sectors}    
}                     
\end{table} 

\newpage

\section{Signal-to-noise ratio and comparison with MLNSL}\label{ap:snr}

\indent We compare how well the MLSS model fits the data with respect to the MLNSL model using the likelihood ratio test. Since the MLNSL model is nested into the MLSS model, we use the $\chi^2$ distribution to test the null hypothesis (the MLSS model does not fit the data better than the MLNSL) against the alternative hypothesis (the MLSS model fits the data better than the MLNSL). The null hypothesis is rejected with a p-value smaller than $0.01$ for both news and social sentiment.

In the last columns of Table \ref{table:static_news_sentiment} in the paper we report the signal-to-noise ratio for each asset obtained using the MLSS model and the signal-to-noise ratio obtained using the MLNSL model. The signal-to-noise ratio for the MLSS model, using the same notation of equation \eqref{eq:sentiment_model}, is evaluated as
\begin{equation}\label{eq:stn_MLSS}
\textsf{stn}(i)^{\text{MLSS}} = \frac{\textsf{Var}\left(\Lambda(i,\cdot) v_t \right) +
\textsf{Var}\left(u^{i}_t \right) }{\textsf{Var}\left(\epsilon^{i}_t \right) } = \frac{\sum_{j=1}^{q}(\Lambda(i,j))^2 + Q_{\text{short}}(i,i) }{R(i,i)}
\end{equation}
while the signal to noise ratio for the MLNSL model, using the notation of equation \eqref{eq:MLNSL}, is evaluated as
\begin{equation}\label{eq:stn_MLNSL}
\textsf{stn}(i)^{\text{MLNSL}} = \frac{\textsf{Var}\left(v^i_t \right) }{\textsf{Var}\left(\epsilon^{i}_t \right) } = \frac{ Q(i,i) }{R(i,i)}
\end{equation}
When the MLSS model is estimated, the signal to noise ratio is on average around $0.8$ for the news sentiment, while when the MLNSL model is estimated, the signal to noise ratio decreases to an average of $0.03$. 

\indent Thus our proposed MLSS model has a signal-to-noise ratio approximately twenty times larger than the MLNSL. Our result also points out that the noise in social media is generally higher than the noise in newspapers.

\newpage

\section{Quantile regression: Contemporaneous effects}\label{ap:Contemporaneous_quantile}
In this appendix we perform the same tests of Section \ref{sec:Lagged_quantile} in the paper with contemporaneous return and sentiment. In particular, we compute the quantile regression \eqref{eq:functional_quantile_reg} with $h = 0$,
\begin{table}[h!]                                                                                 
\centering          
{\small                                                                          
\begin{tabular}{clllll}
\hline                                                                                        
\multirow{2}{*}{\bf{$\tau$ quantiles}} & \multicolumn{5}{c}{\bf{$R^1(\tau)$ measure}}  \\
 & MLSS & LSS & MLNSL & LNSL & Obs \\                                                             
\hline                                                                                        
$0.01$ & $16.2 \%^{***}$ & $6.1 \%^{**}$ & $1.4 \%$ & $1.6 \%$ & $0.6 \%$ \\                  
$0.05$ & $9.2 \%^{***}$ & $4.0 \%^{***}$ & $2.8 \%^{***}$ & $2.7 \%^{***}$ & $1.7 \%^{***}$ \\
$0.10$ & $7.1 \%^{***}$ & $4.3 \%^{***}$ & $3.5 \%^{***}$ & $3.2 \%^{***}$ & $2.5 \%^{***}$ \\
$0.33$ & $2.2 \%^{***}$ & $1.8 \%^{***}$ & $1.9 \%^{***}$ & $1.7 \%^{***}$ & $1.0 \%^{***}$ \\
$0.50$ & $1.1 \%^{***}$ & $1.1 \%^{***}$ & $1.2 \%^{***}$ & $1.0 \%^{***}$ & $0.7 \%^{***}$ \\
$0.66$ & $0.5 \%^{***}$ & $0.9 \%^{***}$ & $1.3 \%^{***}$ & $0.8 \%^{***}$ & $0.7 \%^{***}$ \\
$0.90$ & $1.2 \%^{***}$ & $1.7 \%^{***}$ & $1.5 \%^{***}$ & $0.8 \%^{***}$ & $0.8 \%^{***}$ \\
$0.95$ & $2.9 \%^{***}$ & $2.3 \%^{***}$ & $1.9 \%^{***}$ & $1.0 \%^{***}$ & $1.0 \%^{***}$ \\
$0.99$ & $10.2 \%^{***}$ & $4.6 \%^{**}$ & $0.9 \%$ & $0.6 \%$ & $1.5 \%$ \\   
\hline                                                                                        
\end{tabular}                                                                                 
\caption{\small The $R^1$ measure across the value $\tau$. We denote with $^{***}$ the significance at $1\%$, $^{**}$ the significance at $5\%$ and $^{*}$ the significance at $10\%$ }                                                                      
\label{table:R1_quantile_reg}       
}                                                             
\end{table}
Table \ref{table:R1_quantile_reg} shows the values of the $R^1(\tau)$ measure for different values of $\tau$. It is worth to notice that the quantile regressions are highly significant for every model, except for the $0.01$ and $0.99$ quantiles, where they are only significant for the MLSS and LSS models.
There are three important findings. The first one is that, as in the lagged relation, for any model, the values of $R^1$ are higher in the tails and lower close to the median. The results are not symmetric around the median. The lower quantiles, which correspond to negative returns, have higher $R^1$ than the corresponding $R^1$ in the higher quantiles. This suggests that the sentiment series are powerful explanatory variables in bad times. This conclusion is in accordance with the results in \citep{Garcia2013}, which shows that investors' sensitivity to news is most pronounced going through hard times. The second result is that the models which exploit the multivariate structure (MLSS and MLNSL) produce higher $R^1$ measures than the corresponding models which apply the cross-sectional averaging procedure on the sentiment series (LSS and LNSL models, respectively). This result confirms that the cross-sectional dependence structure is helpful in extracting a sensible signal. The last result is that the MLSS and LSS models, excluding few values around the median, have higher $R^1(\tau)$ values than other models. This suggests that disentangling the long-term and short-term sentiment components is the most important step to capture the contemporaneous relation with market returns. In particular, the MLSS model, which exploits both the separation in two components and the multivariate structure, strongly outperforms the benchmark model, which solely uses the observed noisy sentiment.

If we look at the contribution of the short and long-term sentiment separately using equation \eqref{eq:test_quantile_reg}, we again observe similar results with the one observed in Section \ref{sec:Lagged_quantile} of the paper.
\begin{table}                       
\centering     
{\small                     
\begin{tabular}{cll}            
\hline          
\multirow{2}{*}{\bf{$\tau$ quantiles}} & \multicolumn{2}{c}{\bf{$p$-values}}             \\       
 & $L_t^{ST}$ &  $L_t^{LT}$ \\             
\hline                              
$0.01$ & $0.005 \%$ & $76.313 \%$ \\
$0.05$ & $0.000 \%$ & $2.052 \%$ \\
$0.10$ & $0.000 \%$ & $3.381 \%$ \\
$0.33$ & $0.000 \%$ & $3.257 \%$ \\
$0.50$ & $0.000 \%$ & $7.668 \%$ \\
$0.66$ & $1.487 \%$ & $20.078 \%$ \\
$0.90$ & $0.189 \%$ & $0.309 \%$ \\
$0.95$ & $0.007 \%$ & $0.922 \%$ \\
$0.99$ & $0.006 \%$ & $22.903 \%$ \\
\hline                              
\end{tabular}                       
\caption{\small p-values for the statistics in equation \eqref{eq:statistic_ST_quant_reg} and equation \eqref{eq:statistic_LT_quant_reg}.}
\label{table:p_val_MLSS_quant_reg}   
}       
\end{table}                         
Table \ref{table:p_val_MLSS_quant_reg} reports the p-values of the statistics \eqref{eq:statistic_ST_quant_reg} and \eqref{eq:statistic_LT_quant_reg} and shows that the short-term sentiment is highly significant at any level of $\tau$, while the long-term sentiment has lower p-values. In particular, the short-term sentiment, which captures rapidly changing trends, is significant for extreme returns ($\tau = 0.01$ or $\tau = 0.99$) while the long-term sentiment is not. This result suggests that extreme market swings can be explained by unexpected and short-lasting news. Moreover, it further supports the importance of disentangling sentiment components which are sensitive to different time scales.\\
\indent These findings show very strong contemporaneous relation between sentiment and market returns. We look at these results as a sanity check of our approach. Indeed, since we are not claiming that sentiment causes returns or viceversa, then it is reasonable to expect a significant contemporaneous relation at daily time scale. The sentiment explains returns and this could  be due to the fact that the news, from which sentiment is computed, report and comment about the market performance. What is more promising is that the $R^1$ measure increases with the complexity of the model, and this is especially true for extreme market events -- where the observed sentiment is not significant. Then, we conclude that an essential ingredient of the analysis is the combination of a multivariate model with the separation of sentiment in two components, the stochastic long-run trend (long-term sentiment) common to all assets and a fast changing and asset-specific trend (short-term sentiment).

\newpage

\section{Portfolio allocation on February 2007-June 2017 without trading costs}\label{ap:without_trading_costs}
\begin{figure}[htbp]
\hspace*{-1.0cm}
\begin{center}
\includegraphics[scale=0.3]{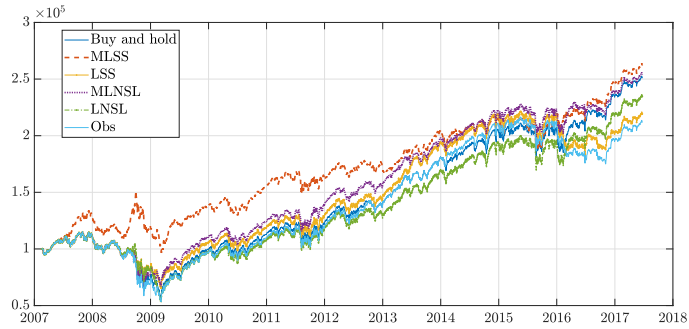} 
\caption{\small Portfolio evolution of the sentiment based strategies built using equation \eqref{eq:definition_portfolio_costs} together with the buy-and-hold equally weighted portfolio in blue.} \label{fig:market_portfolio_bh_short_selling_no_cost}
\end{center}
\end{figure}

\indent Table \ref{table:portfolio_market_short_selling_no_cost} reports the performances of the five sentiment strategies together with the buy-and-hold portfolio without trading costs. We notice that the qualitative results do not change. Given that the MLSS portfolio produces the higher number of trades, the performance gap with respect to the other strategies increase in size in terms of returns, Sharpe and Sortino ratios. Figure \ref{fig:market_portfolio_bh_short_selling_no_cost} shows the evolution of the sentiment-based portfolios without trading costs. The MLSS portfolio, contrary to all the other portfolios, performs very well during the financial crisis and strongly outperforms the other sentiment-based portfolios and the $1/n$ portfolio. However, the gain reduces during the $2009-2017$ bull market period. Nonetheless, even if the absolute return reduces, the volatility is consistently lower during the whole period and the Sharpe and Sortino ratios are respectively $40\%$ and $49\%$ higher in the MLSS portfolio rather than the buy-and-hold portfolio.
\begin{table}[h!]
\centering
{\small
\begin{tabular}{c|llllll}
\hline
Measures & BH & MLSS & LSS & MLNSL & LNSL & Obs \\
\hline
A. return (\%) & $8.972 $ & \textbf{9.393} & $7.650 $ & $9.091 $ & $8.308 $ & $7.33 $ \\                    
A. volatility (\%) & $19.122 $ & \textbf{14.080} & $17.996 $ & $17.797 $ & $19.113 $ & $19.765 $ \\          
A. neg. volatility (\%) & $15.514 $ & \textbf{10.932} & $14.294 $ & $14.368 $ & $15.125 $ & $15.928 $ \\
A. Sharpe ratio & $0.469 $ & \textbf{0.667} & $0.425 $ & $0.511 $ & $0.435 $ & $0.371 $ \\                         
A. Sortino ratio & $0.578 $ & \textbf{0.859} & $0.535 $ & $0.633 $ & $0.549 $ & $0.46 $ \\                         
MDD (\$) & $59377 $ & $54938 $ & $50182 $ & \textbf{49397} & $61921 $ & $61982 $ \\                    
\hline
\end{tabular}
\caption{\small Performances of the six strategies without trading cost for the period February $2007$ - June $2017$. In bold, the best performance per row. BH is the buy-and-hold portfolio, while MLSS, LSS, MLNSL, LSNSL, and Obs correspond to portfolios built from the corresponding model for the sentiment time series.}
\label{table:portfolio_market_short_selling_no_cost}
}
\end{table}

\newpage

\section{Robustness check: Portfolio allocation on July 2017 - December 2019}\label{ap:robustness_check}
In this appendix, we use the parameter values estimated from the TRMI sentiment time series over the February $2007$ - June $2017$ to filter the sentiment signal in the July $2017$ - December $2019$ period. This procedure ensures that the filtered signals do not suffer from any forward-looking bias.

Table \ref{table:portfolio_market_short_selling_no_cost_period2} shows that the qualitative results do not change from the Section \ref{sec:2007_2017}. The MLSS model outperforms the buy-and-hold portfolio with a relative gain of $14\%$ in both Sharpe and Sortino ratio. Two main differences are visible from the February $2007$ - June $2017$ period. The LSS model slightly outperforms the buy-and-hold portfolio and it is the second best performing model, while in the previous case the second best performing model was the MLNSL. The Obs portfolio produces the same performance of the buy-and-hold portfolio and the reason is that the selling signal from $s^{\text{Obs}}$ is always negative. Then, the number of transaction is equal to $1$. In table \ref{table:portfolio_market_short_selling_cost_period2}, we see that the transaction costs do not change the qualitative results and again, the MLSS strategy is the one which produces the higher number of trades and, as a consequence, the higher transaction costs.
As done in the main text, the performance of the MLSS strategy from table \ref{table:portfolio_market_short_selling_cost_period2} can be further improved by applying the penalization of the selling signal based on the Mc Fadden's $R^2$.

\begin{table}                                                                                                    
\centering                                                                                                       
\begin{tabular}{c|llllll}                                                                                 
\hline                                                                                                           
Measures & BH & MLSS & LSS & MLNSL & LNSL & Obs \\                                                                      
\hline                                                                                                           
A. return (\%) & $13.56 $ & \textbf{15.28} & $14.316 $ & $12.584 $ & $9.72 $ & $13.56 $ \\                 
A. volatility (\%) & $14.477 $ & \textbf{14.311} & $14.257 $ & $14.672 $ & $15.053 $ & $14.477 $ \\         
A. neg. volatility (\%) & $10.673 $ & $10.552 $ & \textbf{10.489} & $10.841 $ & $11.521 $ & $10.673 $ \\
A. Sharpe ratio & $0.937 $ & \textbf{1.068} & $1.004 $ & $0.858 $ & $0.646 $ & $0.937 $ \\                  
A. Sortino ratio & $1.27 $ & \textbf{1.448} & $1.365 $ & $1.161 $ & $0.844 $ & $1.27 $ \\                   
MDD (\$) & $23489 $ & \textbf{18871} & $19631 $ & $26639 $ & $31207 $ & $23489 $ \\ 
\hline                                                                                                           
\end{tabular}                                                                                                    
\caption{Performances of the six strategies without trading cost for the period July $2017$ - December $2019$. In bold, the best performance per row. BH is the buy-and-hold portfolio, while MLSS, LSS, MLNSL, LSNSL, and Obs correspond to portfolios built from the corresponding model for the sentiment time series.}
\label{table:portfolio_market_short_selling_no_cost_period2}
\end{table}

\begin{table}
\centering
\begin{tabular}{c|llllll}
\hline
Measures & BH & MLSS & LSS & MLNSL & LNSL & Obs \\
\hline
A. return (\%) & $13.566 $ & \textbf{14.571} & $13.907 $ & $12.518 $ & $9.492 $ & $13.566 $ \\             
A. volatility (\%) & $14.483 $ & $14.394 $ & \textbf{14.328} & $14.7 $ & $15.122 $ & $14.483 $ \\           
A. neg. volatility (\%) & $10.678 $ & $10.641 $ & \textbf{10.564} & $10.865 $ & $11.589 $ & $10.678 $ \\
A. Sharpe ratio & $0.937 $ & \textbf{1.012} & $0.971 $ & $0.852 $ & $0.628 $ & $0.937 $ \\                  
A. Sortino ratio & $1.27 $ & \textbf{1.369} & $1.317 $ & $1.152 $ & $0.819 $ & $1.27 $ \\                   
MDD (\$) & $23489 $ & \textbf{19319} & $20330 $ & $26868 $ & $31898 $ & $23489 $ \\                    
Number of trades & $1 $ & $41 $ & $25 $ & $5 $ & $13 $ & $1 $ \\
Transaction costs (\$) & $50 $ & $2460 $ & $1421 $ & $279 $ & $741 $ & $50 $ \\
\hline
\end{tabular}
\caption{Performances of the six strategies with trading cost for the period July $2017$ - December $2019$. In bold, the best performance per row. BH is the buy-and-hold portfolio, while MLSS, LSS, MLNSL, LSNSL, and Obs correspond to portfolios built from the corresponding model for the sentiment time series.}
\label{table:portfolio_market_short_selling_cost_period2}
\end{table}

\newpage

\section{Statistical significance of the sentiment portfolios}\label{ap:portfolio_significance}
Here, we assess the significance of the trades produced by the strategy \eqref{eq:trading_signal_short_selling} for the different sentiment filters. We design a Monte Carlo experiment where a trader follow a random selling signal. The selling signal is given by $s^{\text{shuffled, mod}}_t $, which is nothing more than a shuffled realization of $s^{\text{mod}}_t$. The number of random selling signals corresponds by construction to the number of selling signals produced by $s^{\text{mod}}$, which is reported in table \ref{table:portfolio_market_short_selling_cost} of the paper.  We repeat the experiment $10,000$ times. The corresponding portfolios are then sorted according to their Sharpe and Sortino ratios and the p-value of each strategy is computed by comparison with the quantiles from the Monte Carlo experiment. Table \ref{table:app:portfolio_market_simulations} shows the results over the period February $2007$ - June $2017$. The MLSS strategy significantly outperforms the random strategy with a p-value smaller than $5\%$. All the other strategies are not statistically different from a random strategy. 

\begin{table}                                                                                                    
\centering                                                                                                       
\begin{tabular}{cll}                                                                                 
\hline                                                                                                           
Strategies & Annual Sharpe ratio & Annual Sortino ratio \\
\hline                                                                                                           
MLSS & $0.519 (96.9\%)$ & $0.725 (96.4\%)$ \\
best $5 \% $ & $0.485$ & $0.69$ \\                        
best $10 \% $ & $0.431$ & $0.611$ \\                      
best $25 \% $ & $0.349$ & $0.491$ \\                      
median & $0.26$ & $0.364$ \\                              
\hline
LSS & $0.385 (36.7\%)$ & $0.542 (37.4\%)$ \\
best $5 \% $ & $0.529$ & $0.749$ \\                      
best $10 \% $ & $0.502$ & $0.709$ \\                     
best $25 \% $ & $0.457$ & $0.644$ \\                     
median & $0.409$ & $0.574$ \\                            
\hline         
MLNSL & $0.495 (86.2\%)$ & $0.7 (86.9\%)$ \\
best $5 \% $ & $0.523$ & $0.739$ \\                    
best $10 \% $ & $0.504$ & $0.711$ \\                   
best $25 \% $ & $0.474$ & $0.667$ \\                   
median & $0.443$ & $0.622$ \\
\hline
LNSL & $0.419 (27.0\%)$ & $0.59 (27.6\%)$ \\
best $5 \% $ & $0.524$ & $0.741$ \\                    
best $10 \% $ & $0.505$ & $0.713$ \\                   
best $25 \% $ & $0.475$ & $0.669$ \\                   
median & $0.446$ & $0.627$ \\                           
\hline                                   
\end{tabular}                                                                                                    
\caption{Performances of the sentiment strategies compared with the $95\%$, $90\%$, $75\%$ and $50\%$ quantiles from the shuffled strategy. Values in brackets are the percentages of randomly generated portfolios which perform worse than the sentiment-based strategy for the period February $2007$ - June $2017$.}                                                                                         
\label{table:app:portfolio_market_simulations}
\end{table}                                                                                                      
The p-values of the MLSS trading strategy are even lower when the $R^2$-penalized trading strategy \eqref{eq:R_squared_correction2} is implemented. Table \ref{table:app:portfolio_market_simulations_R2} shows the p-values. When the number $(100\%)$ is reported, all the $10,000$ random strategies perform worse than the MLSS $\alpha$ strategy. The number of selling signals for the MLSS with $\alpha=0.80$ is too small and the result may be not reliable.

\begin{table}                                                                                                    
\centering                                                                                                       
\begin{tabular}{cll}                                                                                 
\hline                                                                                                           
Strategies & Annual Sharpe ratio & Annual Sortino ratio \\
\hline                                                                                                           
MLSS(00) & $0.519 (96.9\%)$ & $0.725 (96.4\%)$ \\
best $5 \% $ & $0.485$ & $0.69$ \\                        
best $10 \% $ & $0.431$ & $0.611$ \\                      
best $25 \% $ & $0.349$ & $0.491$ \\                      
median & $0.26$ & $0.364$ \\                              
\hline
MLSS(20) & $0.525 (96.5\%)$ & $0.735 (96.1\%)$ \\
best $5 \% $ & $0.502$ & $0.712$ \\                       
best $10 \% $ & $0.457$ & $0.648$ \\                      
best $25 \% $ & $0.383$ & $0.539$ \\                      
median & $0.306$ & $0.428$ \\                             
\hline         
MLSS(35) & $0.68 (99.9\%)$ & $0.967 (99.8\%)$ \\
best $5 \% $ & $0.512$ & $0.727$ \\                      
best $10 \% $ & $0.473$ & $0.67$ \\                      
best $25 \% $ & $0.409$ & $0.576$ \\                     
median & $0.339$ & $0.475$ \\  
\hline
MLSS(50) & $0.723 (100.0\%)$ & $1.03 (100.0\%)$ \\
best $5 \% $ & $0.531$ & $0.752$ \\                        
best $10 \% $ & $0.495$ & $0.7$ \\                         
best $25 \% $ & $0.436$ & $0.615$ \\                       
median & $0.373$ & $0.523$ \\ 
\hline
MLSS(65) & $0.757 (100.0\%)$ & $1.09 (100.0\%)$ \\
best $5 \% $ & $0.534$ & $0.757$ \\                        
best $10 \% $ & $0.504$ & $0.711$ \\                       
best $25 \% $ & $0.459$ & $0.647$ \\                       
median & $0.415$ & $0.583$ \\  
\hline
MLSS(80) & $0.534 (97.3\%)$ & $0.752 (97.2\%)$ \\
best $5 \% $ & $0.519$ & $0.733$ \\                       
best $10 \% $ & $0.501$ & $0.707$ \\                      
best $25 \% $ & $0.477$ & $0.671$ \\                      
median & $0.45$ & $0.634$ \\                            
\hline                                   
\end{tabular}                                                                                                    
\caption{Performances of the $R^2$ adjusted MLSS strategies for different values of $\alpha$ compared with the $95\%$, $90\%$, $75\%$ and $50\%$ quantiles from the random strategy for the period February $2007$ - June $2017$. The sentiment strategies are referred to as MLSS($\alpha\times 100$). Values in brackets are the percentages of randomly generated portfolios which perform worse than the sentiment-based strategy. }                                                                                         
\label{table:app:portfolio_market_simulations_R2}
\end{table}                                                                                                      

\newpage

\end{document}